\documentclass{aa}  

\usepackage{graphicx}
\usepackage{txfonts}
\begin{document}

   \title{Updated properties of the old open cluster Melotte~66: Searching for multiple stellar populations\thanks{Based on observations   collected at Paranal  Observatory under program 088.D-0045 and 076.D-0220, and at Las Campanas Observatory}}

\author{Giovanni Carraro\thanks{On leave from Dipartimento di Fisica e Astronomia, Universit\'a di Padova, Italy} 
          \inst{1}, Gayandhi de Silva\inst{2}, Lorenzo Monaco\inst{1}, Antonino P. Milone\inst{3}, Renee Mateluna\inst{4}
          }

     \institute{ESO, Alonso de Cordova 3107, 19001,
           Santiago de Chile, Chile\\
              \email{gcarraro,lmonaco@eso.org}
         \and
         Australian Astronomical Observatory, 105 Delhi Rd, NSW 2113, Australia;
         \email{gdesilva@aao.gov.au}
                  \and
             Research School of Astronomy and Astrophysics, Australian National University, Mt Stromlo Observatory, Cotter Rd, Weston, ACT 2611, Australia\ 
         \email{antonino.milone@anu.edu.au}
          \and                  
            Departamento de Astron\'onia, Universidad de Concepci\'on,
               Casilla 169, Concepcion, Chile\\
               \email{rmateluna@astro-udec.cl}
             }

   \date{Received ....; accepted...}

 
  \abstract
   {Multiple generations of stars are routinely encountered in globular clusters but no convincing evidence has been found in Galactic open clusters  to date.}   
   {In this paper we use new photometric and spectroscopic data to search for multiple stellar population signatures in the old, massive open cluster, Melotte~66. The cluster is known
   to have a red giant branch wide in color, which could  be an indication of metallicity spread. Also the main sequence is wider
   than what is expected from photometric errors only. 
   This evidence might be associated  with either 
   differential reddening or binaries. Both hypothesis  
    have, however, to be evaluated in detail before
   recurring to  the presence of multiple stellar populations.}
   {New, high-quality, CCD UBVI photometry have been acquired to this aim with high-resolution spectroscopy of seven clump stars, that are complemented with literature data; this  doubles the number of clump star member of the cluster for which high-resolution spectroscopy is available.
   All this new material is carefully analyzed in search for any spectroscopic or photometric manifestation of multiple populations among the cluster stars.}
   {Our photometric study confirms that the width of the main sequence close to the turn off point is entirely accounted for by binary stars and differential reddening, with no need to advocate more sofisticated scenarios, such as metallicity spread
   or multiple main sequences. By constructing synthetic color-magnitude diagrams, we infer that the binary fraction has to be as large as 30$\%$ and their mass ratio in the range 0.6-1.0. As a by-product of our
   simulations, we provide new estimates of the cluster fundamental parameters. We measure a reddening E(B-V)=0.15$\pm$0.02, and confirm the presence of a marginal differential reddening. The distance to the cluster
   is $4.7^{+0.2}_{-0.1} $kpc and the age is 3.4$\pm$0.3 Gyr, which is  somewhat younger and better constrained than previous estimates.}
   {Our detailed abundance analysis reveals that, overall, Melotte~66 looks like a typical object of the old thin disk population with no significant spread in any of the  chemical species we could measure.
       Finally, we perform a photometric study of the blue straggler star population and argue that their number in Melotte~66 has been significantly overestimated in the past. The analysis of their spatial distribution supports the scenario that they are most probably primordial binaries.}


\authorrunning{Carraro et al.}
\titlerunning{The old open cluster Melotte 66}

   \maketitle
%

\section{Introduction}
With the exception of Ruprecht~106 (Villanova et al. 2013) and possibly Terzan~8 (Carretta et al. 2014), all the Milky Way 
old globular clusters studied so far show  either photometric or spectroscopic signatures  of multiple stellar populations.
The parameter driving the presence or absence of more than one population seems to be the total mass, and much work
has currently been done to study the lowest  mass globulars. As stressed by Villanova et al. (2013), Terzan~7, Palomar~3, and 
NGC~1783 can be good candidates to look at.\\

\noindent
An interesting and different perspective can be to consider  Galactic open clusters -  in particular those few old open clusters
that are still  massive enough - and search for a signature of multiple generations among their stars.
Unfortunately, old massive, open clusters are extremely rare in the Milky Way:  first, because open clusters are not very massive at birth  and second,  because they loose quite some mass during their lifetime, mostly due to the tidal interaction with the Milky Way
and the dense environment of the Milky Way disk (Friel 1995).
The potential interest of old open clusters in the context of multiple stellar generations has been recognized for a while,
but so far only two clusters have been investigated in details: NGC~6791 (Geisler et al. 2012) and Berkeley~39 
(Bragaglia et al. 2012). Both clusters have current masses $\sim 10^4 M_{\odot}$; NGC~6791  is probably  somewhat more
massive than Berkeley~39. 
In the case of Berkeley~ 39,  no signature of multiple populations were found, while there seem to be two groups of stars in NGC~6791 having different Na abundance.\\

\noindent
It is important, however, to state as clear as possible that masses are difficult to estimate for open clusters because of the significant field star contaminations and the large presence of binaries, which affects both photometric and kinematic mass measures (Friel 1995).\\
Therefore, one is often left with crude mass estimates which are  based mostly on the appearance of the color magnitude diagram (CMD)
and the number of, for example, clump stars. 
A visual inspection at old open cluster older than, say, 5 Gyrs, shows that we are left with maybe only  three probably massive star clusters besides NGC6791 and Berkeley~39.  
They are Trumpler~5, Collinder~26,1 and Melotte~66. No estimate of their mass is available, but
a quick inspection of their CMD immediately shows that they harbor roughly the same number of clump stars as NGC~6791 and 
Berkeley~39, and therefore, their mass should be roughly of the same order.
It seems to us therefore urgent to look at these few clusters, and in this paper we are going to discuss new photometric and spectroscopic material for one of them: Melotte 66.\\

\noindent
The plan of the paper is as follows. In Section~2, we summarize  the literature information on Melotte~66 as completely as possible. Section~3 describes our photometric dataset,
and provides details on observation,  data reduction, and standardization.  A star count analysis is then performed in Section~4. Section~5 deals with the study of Melotte~66 photometry, and the derivation of its fundamental parameters via the comparison with theoretical models. In Section~6, we describe the spectroscopic data, while  we perform a detailed abundance analysis in Section~7 .  The blue straggler population in Melotte~66 is investigated in Section~8 and, finally, the conclusions of our work are drawn in Section~9.

  \begin{figure}
   \centering
   \includegraphics[width=\columnwidth]{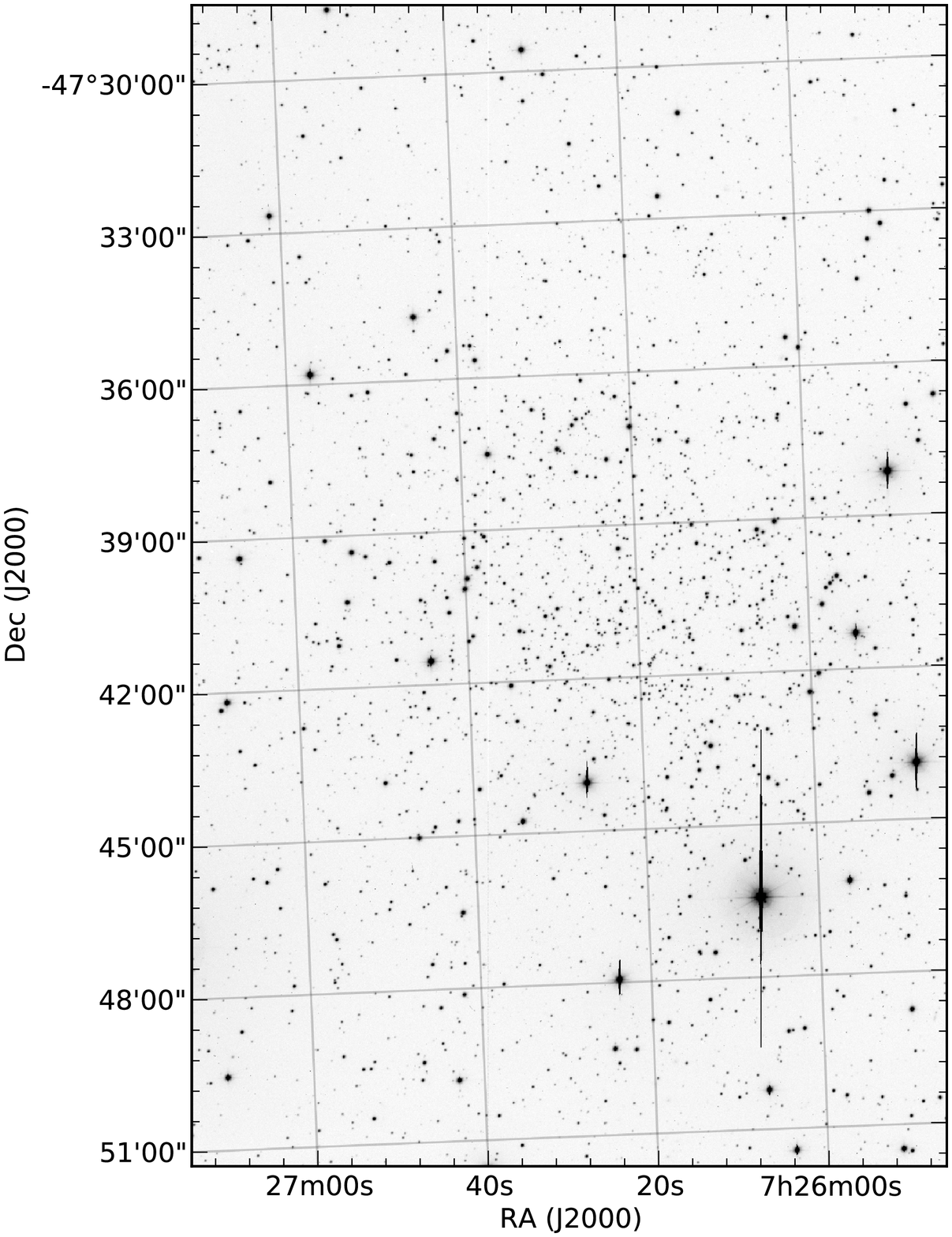}
   \caption{An example of a CCD frame centered on Melotte~66. North is up, east to the left, and the field
   of view is 14.8 $\times$ 22.8 arcmin. The image is in the B filter, and the exposure was 1500 secs.}
    \end{figure}

\section{Melotte 66 in the literature}
Melotte~66 has been the subject of many studies over the years.
It was recognized early on as a potentially old cluster by  King (1964) and  Eggen \& Stoy (1962) but studied
in details for the first time only ten  years later by Hawarden (1976, 1978). In these two papers, Hawarden highlighted very clearly
what makes Melotte~66 particularly interesting. He made use of UBV photometry which is  mostly photographic,
to derive a color-magnitude diagram (CMD) that revealed the cluster turn off (TO) region for the first itme.  \\
Based on this
diagram, Hawarden draw the attention on (1) a rich and  wide-in-color  red giant branch (RGB),  (2) the lack of a subgiant region, (3)
a prominent gap in the upper main sequence, (4) the presence of a group of blue stragglers, (5)
a  conspicuous clump of core He- burning stars and, finally, (6) an anomalous low metallicity ([Fe/H]$\sim$-0.3), as inferred from the 
cluster ultraviolet excess index $\delta(U-B)_{0.6}$ .
He also provided estimates of the cluster absolute distance modulus (12.4 mag),  reddening (0.17), and age ($\sim$ 6-7 Gyr),
emphasizing its extreme location below the Galactic plane ($\sim$ 750 pc).\\

\noindent
These findings boosted a wealth of investigations aimed at understanding the cluster peculiarities, especially the low metallicity and the color spread at fixed luminosity in the RGB.
Hawarden (1978)added a few more photoelectric observations and concluded that differential reddening
is not the culprit for the RGB width in a short contribution  . \\
Anthony-Twarog et al. (1979) improved the BV photometry of the cluster and provided a CMD of higher quality than Hawarden,
although with about the same magnitude limits. They also provided low resolution spectra  and 
the star distribution in the CMD with theoretical isochrones  for the first time. Their study reproduced all the features found by Hawarden, confirmed
the metal abundance derived from his work on UBV photometry and from Dawson (1978) DDO\footnote{David Dunlap Observatory photometric system} photometry, and provided further
support to Melotte 66's old age (6-7 Gyrs), which about 1 Gyr older than NGC~188, and makes Melotte~66 the oldest open cluster know
in those years. Finally, based on their CN strength, some  blue RGB
were suggested to be asymptotic giant branch (AGB) stars.\\

\noindent
The first modern CCD study was conducted by Kaluzny and Shara (1988) in the BV pass-bands in a search for contact binaries.
For the very first time,  the MS of the cluster was revealed down to V $\sim$ 20. The MS looked quite wide in perfect similarity to several Large Magellanic Cloud (LMC) clusters (Milone et al. 2009).
No clear evidence of a significant binary population was, nonetheless, found.\\

\noindent
A crucial step ahead in our understanding of these cluster properties come with the study by Anthony-Twarog et al. (1994). They
presented CCD ubyH$\beta$ photometry down to V $\sim$ 20. The MS was found to be much wider than expected from photometric errors, and the RGB width smaller than what expected from the MS. The subgiant branch was confirmed to be poorly populated.
A metallicity spread implied by {\it m1}\footnote{$m1 = (v-b)-(b-y)$ is an index in the Stromgren system sensitive to metallicity
(Crawford  1958).} index was found among RGB stars.  Differential reddening across the cluster surface was excluded as the cause for the color spread in the CMD. The metallicity variation was also deemed to be not enough to account for the MS broadness, while star-to-star CN variation was conceived to be the most viable explanation for the color variation among RGB stars.\\

\noindent The photometric study that  followed (Kassis et al. 1997, Zloczewski et al. 2007) provided very deep CCD photometry, extending the color coverage to the I band  and  revealed   a clear binary  sequence parallel to the star cluster MS for the first time.  Zloczeskwi et al.  also constructed a reddening map, showing that reddening variations are significant across the cluster surface , at
odds with Anthony-Twarog et al. ( 1994)  study.\\

\noindent
Finally, important pieces of information come from the spectroscopic studies carried out in the meanwhile.
Friel \& Janes (1993) and Friel et al. (2002) first obtained medium resolution spectra of four giants in Melotte 66 and measured [Fe/H]=-0.51$\pm$0.11,
confirming that the cluster is indeed one of the most metal poor old open clusters in the Milky Way
but does not show any significant spread in abundance.\\
Furthermore, two studies presented high resolution spectroscopy:  Gratton \& Contarini (1994) and Sestito et al. (2008). The former
obtained spectra of two giant stars, and concluded that [Fe/H]=-0.38$\pm$0.15, while the latter, a more detailed study, derived
[Fe/H]=-0.33$\pm$0.03 from seven giants.
This figure is very similar to Gratton and Contarini (1994) and confirms that no spread in metallicity was detected among Mel~66 giants.

      \begin{figure}
   \centering
  \includegraphics[width=\columnwidth]{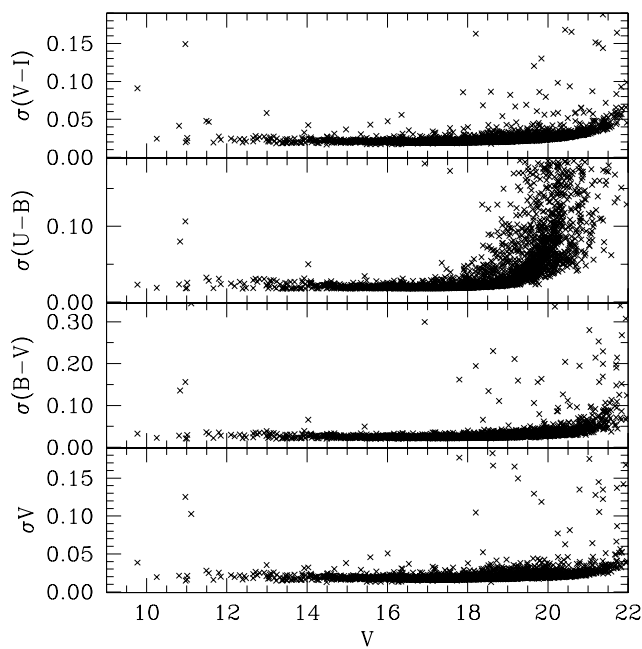}
   \caption{Trend of global photometric errors in magnitude and colors as a function of V magnitude.
   See text for details.}
    \end{figure}

\section{Observations and data reduction: Photometry}

The star cluster Melotte~66  was observed at Las Campanas Observatory (LCO) on the nights from January  3 to January 7,
2011, as illustrated in Table~1, which summarizes useful details of the observations,
like filter coverage, airmass range, exposure time,  and sequences. 
We used the SITe$\#$3 CCD detector  onboard
 the Swope 1.0m telescope\footnote{http://www.lco.cl/telescopes-information/henrietta-swope/}. 
With a pixel scale of 0.435 arcsec/pixel, this CCD allows
to cover 14.8 $\times$ 22.8 arcmin$^2$ on sky. The nights  for which we observed standard stars were photometric with a seeing
range from 0.8 to 1.5 arcsec. The field we covered is shown in Fig.~1, where a bias- and flat-field-
corrected image in the B band (1500 s) is shown.\\
We took  a grand total of 68 images with a small jitter pattern, and  eventually 
the montage frame covered an area of 16.4 $\times$22.8 arcmin$^2$ on sky.\\

\noindent
To determine the transformation from our instrumental system to the standard Johnson-Kron-Cousins
system and to correct for extinction, we observed stars in Landolt's areas
SA~98 (Landolt 1992), that  contains many stars with good absolute photometry and wide color range.
The field was observed
multiple times with different
air-masses ranging from $\sim1.05$ to $\sim1.9$ and covering quite a large color range 
-0.3 $\leq (B-V) \leq$ 1.7 mag.
We
secured night-dependent calibrations (on January 05 and 06), which we then merged, after checking for
stability.

\begin{table}
\tabcolsep 0.1truecm
\caption{$UBVI$ photometric observations of Melotte~66  and the standard star field SA~98.}
\begin{tabular}{lcccc}
\hline
\noalign{\smallskip}
Date & Field & Filter & Exposures (s) & airmass (X)\\
\noalign{\smallskip}
\hline\hline
\noalign{\smallskip}
Jan 03, 2011  & Mel~66  & \textit{U} & 60, 3x300, 1500           &1.07$-$1.12\\
\hline
Jan 04, 2011  & Mel~66 & \textit{U}  & 1500                              & 1.13 \\
                                  &                     & \textit{B}  & 3x45,3x300, 1500       & 1.06$-$1.13\\
\hline
Jan 05, 2011  &Mel~66  & \textit{B}   & 1500                              & 1.13 \\
                                  &                     & \textit{V}  & 30,3x300, 2x1200       & 1.07$-$1.13\\
                                  &                     & \textit{I}    & 30                                   & 1.06\\
                                  & SA 98         & \textit{U}  & 4x240                            & 1.17$-$1.67\\
                                  &                     & \textit{B}  & 4x120                            &  1.18$-$1.55\\
                                  &                     & \textit{V}  &  4x60                             &  1.18$-$1.52\\
                                  &                     & \textit{I}    & 4x60                              &  1.20$-$1.48\\
\hline
Jan 06, 2011  & Mel~66 & \textit{U}   & 5x60, 3x300, 2x1500   & 1.05$-$1.14 \\
                                  &                     & \textit{B}  & 2x45, 300                       & 1.10$-$1.13\\
                                  &                     & \textit{V}  & 2x30, 300                      & 1.09$-$1.13\\
                                  &                     & \textit{I}    & 3x30, 4x300, 2x1000    & 1.08$-$1.30\\
                                  & SA 98         & \textit{U}  & 4x240                            & 1.18$-$1.84\\
                                  &                     & \textit{B}  & 4x120                            &  1.19$-$1.76\\
                                  &                     & \textit{V}  &  4x60                             &  1.20$-$1.70\\
                                  &                     & \textit{I}    & 4x60                              &  1.21$-$1.64\\
\hline
Jan 07, 2011  &  Mel~66 &\textit{U} & 1500                              & 1.42\\ 
                                  &                      &\textit{B}  & 3x45 ,3x300, 2x1500   & 1.15$-$1.30 \\
                                  &                       & \textit{V}  & 2x30, 3x300, 2x1200   & 1.05$-$1.10\\
                                  &                       & \textit{I}  & 3x30, 3x300, 2x1000    & 1.06$-$1.10\\
\hline
\noalign{\smallskip}
\hline
\end{tabular}
\end{table}

\subsection{Basic photometric reduction}
Basic calibration of the CCD frames was done using IRAF\footnote{IRAF is distributed
by the National Optical Astronomy Observatory, which is operated by the Association
of Universities for Research in Astronomy, Inc., under cooperative agreement with
the National Science Foundation.} package CCDRED. For this purpose, zero exposure
frames and twilight sky flats were taken every night.  
All the frames were pre-reduced by applying trimming, bias and flat-field
correction. Before flat-fielding, all frame were corrected for linearity,
following the recipe discussed in Hamuy et al. (2006).\\
Photometry was then performed
using the IRAF DAOPHOT/ALLFRAME and PHOTCAL packages. Instrumental magnitudes were extracted
following the point-spread function (PSF) method (Stetson 1987). A quadratic, spatially-variable
 master PSF (PENNY function) was adopted, because of the large field
of view of the detector. Aperture corrections were then determined,
making aperture photometry for a suitable number (typically 15 to 20) of bright, isolated,
stars in the field. These corrections were found to vary from 0.160 to 0.290 mag, depending
on the filter. The PSF photometry was finally aperture corrected filter-by-filter.

 \begin{figure}
   \centering
   \includegraphics[width=\columnwidth]{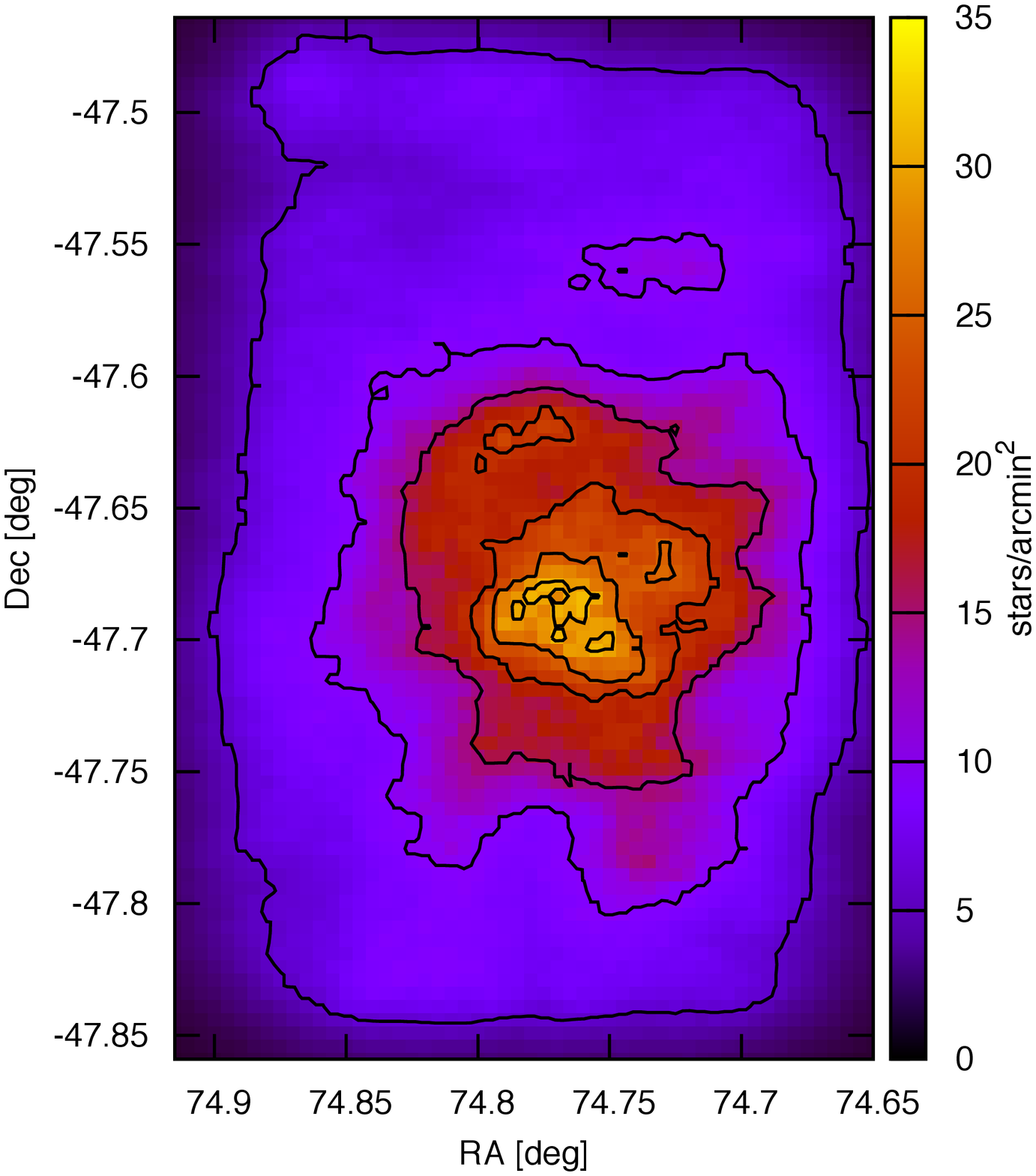}
   \caption{Density contour map for Melotte~66 field. North is up, east to the left, and the field corresponds to 16.4 $\times$22.8 squared arcmin on sky. On the X-axis, RA$\times$ cos($\delta$) is shown
   to keep the same scale as in Fig.~1.}
    \end{figure}
    
   \begin{figure}
   \centering
   \includegraphics[width=\columnwidth]{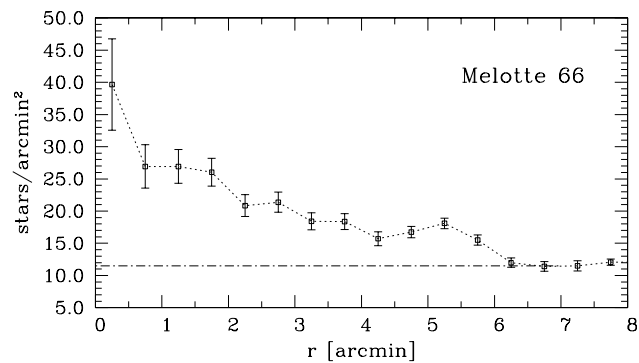}
   \caption{Radial  surface density profile. To define concentric rings, we used the nominal cluster center.}
    \end{figure}

\subsection{Photometric calibration}

After removing problematic stars, and stars having only a few observations in Landolt's
(1992) catalog, our photometric solution  for the run was extracted by combining measures
from both nights after checking if they were stable and similar. This yields
a grand total of 89 measurements per filter and turned out to be:\\

\noindent
$ U = u + (4.921\pm0.012) + (0.47\pm0.01) \times X + (0.079\pm0.014) \times (U-B)$ \\
$ B = b + (3.289\pm0.010) + (0.26\pm0.01) \times X + (0.074\pm0.009) \times (B-V)$ \\
$ V = v + (3.193\pm0.007) + (0.16\pm0.01) \times X - (0.059\pm0.007) \times (B-V)$ \\
$ I = i + (3.505\pm0.009) + (0.08\pm0.01) \times X + (0.052\pm0.006) \times (V-I)$ \\

\noindent
where $X$ indicates the airmass.\\
The final {\it r.m.s} of the fitting in this case was 0.025, 0.018, 0.010, and 0.010 
in $U$, $B$, $V$ and $I$, respectively.\\

\noindent
Global photometric errors were derived  using the scheme developed by Patat \& Carraro
(2001, Appendix A1), which takes the errors resulting from the PSF fitting
procedure (i.e., from ALLSTAR) and the calibration errors (corresponding to the zero point,
color terms, and extinction errors) using errors' propagation into account. In Fig.~2, we present these global photometric errors
in $V$, $(B-V)$, $(U-B)$, and $(V-I)$  plotted as a function of $V$ magnitude. Quick
inspection shows that stars brighter than $V \approx 20$ mag have errors lower than
$\sim0.05$~mag in both magnitude and the $(B-V)$ and $(V-I)$  colors. Larger
errors, as expected, are seen in $(U-B)$.\\

\noindent
The final catalog contains 3474 \textit{UBVI} and 15752 \textit{VI} entries.

\begin{table*}
\tabcolsep 0.1truecm
\caption{An excerpt of the optical photometric table that includes clump with high-resolution spectroscopy
The full version is posted at the CDS website. ID indicates
the numbering. The last four stars have spectra taken from  Sestito et al. (2008).}
\begin{tabular}{cccccccccccc}
ID & Kassis et al. & RA(2000.0) & DEC(2000.0) & V & $\sigma_V$ & (U-B) & $\sigma_{(U-B)}$ & (B-V) & $\sigma_{(B-V)}$ & (V-I) & $\sigma_{(V-I)}$\\    
\hline
 & & deg & deg & & & & & & &\\
\hline
597   & 385   &111.703333 &-47.65939  &14.278 &0.020   &0.802   &0.022   &1.127   &0.028   &1.181   &0.023\\
776   & 603   &111.679583 &-47.61947  &14.378 &0.020   &0.810   &0.022   &1.137   &0.028   &1.142   &0.022\\
1521 &1419 &111.611250 &-47.63216  &14.432 &0.020   &0.747   &0.022   &1.110   &0.028   &1.089   &0.022\\
2099 &1953 &111.572083 &-47.73336  &14.551 &0.021   &0.735   &0.022   &1.118   &0.028   &1.167   &0.023\\
2209 &2155 &111.556250 &-47.62350  &14.454 &0.020   &0.768   &0.022   &1.105   &0.028   &1.108   &0.022\\
2291 &2187 &111.554167 &-47.69819  &14.553 &0.020   &0.746   &0.022   &1.107   &0.028   &1.147   &0.023\\
2803 &2771 &111.510417 &-47.68219  &14.568 &0.020   &0.765   &0.022   &1.117   &0.028   &1.162   &0.023\\
1202 &1000 &111.644045 &-47.71306  &14.640  &0.021  &0.757   &0.022   &1.151   &0.028   &1.182   &0.023\\
1734 &1580 &111.598675 &-47.70005  &14.683 &0.021   &0.762   &0.022   &1.140   &0.028   &1.240   &0.023\\
2980 &2945 &111.491345 &-47.67310  &14.495 &0.020   &0.855   &0.023   &1.139   &0.028   &1.147   &0.022\\
1919 &1805 &111.582833 &-47.67103  &14.665 &0.021   &0.840   &0.023   &1.205   &0.029   &1.195   &0.023\\
\noalign{\smallskip}
\hline
\end{tabular}
\end{table*}

\subsection{Completeness and astrometry}

Completeness corrections were determined by running artificial star experiments
on the data.  Figure~1 clearly shows that Melotte~66 does not suffer from serious crowding,
and therefore, completeness has been evaluated over the whole cluster area.
Basically, we created several artificial images by adding artificial stars
to the original frames, on a frame-by-frame basis. These stars were added at random positions and had the same
color and luminosity distribution of the true sample. To avoid generating overcrowding
we added up to 20\% of the original number of stars in each experiment. Depending on
the frame, between 100 and 500 stars were added. In this way, we have estimated that the
completeness level of our photometry is better than 90\% down to V  $\approx 20.5$ (see Table~3).\\

The optical catalog was then cross-correlated with 2MASS, which resulted in a final catalog that
includes \textit{UBVI} and \textit{JHK$_{s}$} magnitudes. As a by-product, 
pixel (i.e., detector) coordinates
were converted to RA and DEC for J2000.0 equinox, thus providing 2MASS-based astrometry which is  useful
for spectroscopic follow-up.
An excerpt of the optical photometric table used in this investigation is illustrated in Table~2.\\
\noindent

\begin{table}
\caption{Completeness study as a function of the filter.}
\begin{center}
\begin{tabular}{lr  r r r r }
\hline\hline
$\Delta$ Mag   &    U &  B & V &  I\\
\hline
12-13 &   100\%  &   100\%  &  100\%  &   100\% \\ 
13-14 &   100\%  &   100\%  &  100\%  &    100\%  \\ 
14-15 &   100\%  &   100\%  &   100\% &    100\%  \\ 
15-16 &   100\%  &   100\%  &   100\% &    100\% \\ 
16-17 &   100\%  &   100\%  &   100\% &   100\% \\ 
17-18 &   100\%  &   100\%  &   100\% &    100\% \\ 
18-19 &    100\% &   100\%  &    100\% &   100\% \\ 
19-20 &      91\% &     95\%  &     100\% &    100\% \\ 
20-21 &      72\% &     80\%  &      95\% &      96\% \\  
21-22 &      48\&   &    53\%  &      67\% &       88\% \\  
\hline
\end{tabular}
\end{center}
\label{comp}
\end{table}

\subsection{Comparison with previous photometry}
We compared our VI photometry with Kassis et al. (1991), as done by Zloczewski et al. (2007).
From the 2303 common stars we obtain

\begin{equation}
\Delta V =  0.00\pm0.04, and 
\end{equation}

\begin{equation}
\Delta (V-I) = 0.02\pm0.05
\end{equation}

\noindent
in the sense of  subtracting our photometry from the values determined  Kassis et al. (1997). This implies that our VI  photometry is basically in the same system as
Kassis et al. (1997) and Zloczewski et al (2007).\\

\noindent
The comparison in UBV has been done against Zloczewski et al (2007) with the only difference that  the star in common having (U-B) drops to 870 because of the lower sensitivity of these
filters, which are mostly caused by the smaller amount of exposures. We find

\begin{equation}
\Delta (B-V) =  -0.04\pm0.03, and 
\end{equation}

\begin{equation}
\Delta (U-B) = 0.06\pm0.04
\end{equation}

\noindent
We conclude that the two datasets agree fairly well also in UB, and therefore the  two photometries are in the same system.

\section{Star counts and cluster size}
To be able to quantify the amount
of field star contamination, we performed star counts to identify the cluster center and measure its size.
To achieve this, we derived  density contour maps using an array and calculated the density inside each grid step
by a kernel estimate.\\
A quick glance at Fig.~1 shows that Melotte~66 is far from being a symmetric object. 
This is also visible in Fig.~3, which lends further support to the appearance of Fig.~1. 
The cluster is elongated in the direction NE to SW,  and the highest peak does
not represent the center of an uniform star distribution. 
The largest peak is located at RA= 111$^{o}$.57, DEC=-47$^{o}$.71, while the nominal center
of the cluster is clearly displaced to the northeast direction,  at : RA=111$^{o}$.60,  DEC= -47$^{o}$. 66.
One can argue that the most probable reason for such an occurrence is the tidal interaction with the Milky Way. However, we do not have kinematic information, but 
only the cluster radial velocity. High-quality proper motions are still missing, and they would be  very welcome to derive the cluster motion direction and see if this
coincides with the direction of the apparent cluster elongation. \\

\noindent
To isolate probable cluster members,  which are this stars assumed to lie within the cluster boundaries, 
we derive the cluster radial surface density profile, which is shown in Fig.~4.  This has been computed by drawing concentric
rings centered on the nominal cluster center. This is motivated by the fact that the cluster halo still looks almost circular while the densest
central regions look distorted.
Star counts level off at $\sim$ 6 arcmin from the cluster
nominal center, confirming early findings by Hogg\,(1965) that the cluster diameter is around 13 arcmin. The mean density in the field surrounding the cluster is 10 stars/arcmin$^2$ (see also Fig~3), and our survey covers the whole cluster area.
Our estimate of for the radial extent of Melotte~66 is smaller than Zlocewski et al. (2007). This is  most probably because their star counts  beyond 6 arcmin from the cluster center are not properl
area-corrected (concentric rings are not complete anymore), thus producing artificial over densities.
As a consequence, we will adopt a value of  6 arcmin for  cluster radius ion the following.
 We will refer to to the area of the circle with a radius of 6 arcmin as the cluster area, while the area outside 6 arcmin from the cluster center is  referred to as the  offset field.

   \begin{figure*}
   \centering
   \includegraphics[width=15cm, height=13cm]{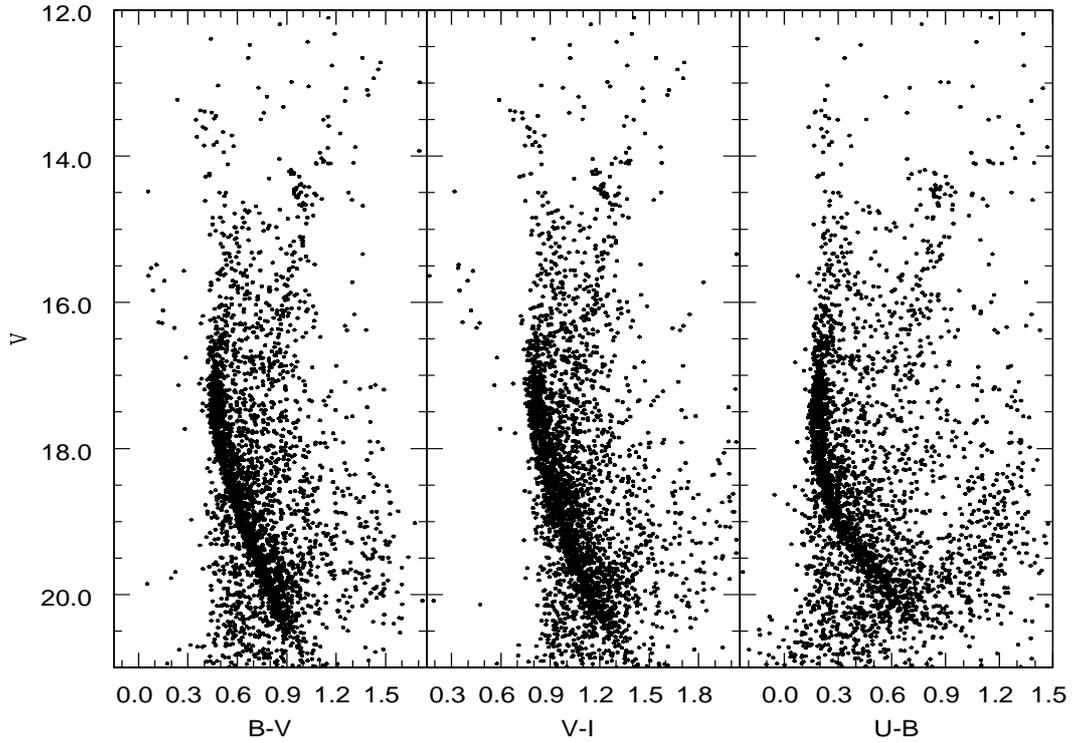}
   \caption{CMD of Melotte~66. All measured stars are shown. }
    \end{figure*}

\section{Photometric diagrams}
The CMDs for all the stars in the observed field are shown in Fig~4 for three different color combinations: V/B-V, V/V-I, and V/U-B
from the left to the right.
All the features previously outlined by Anthony-Twarog et al. (1994) are visible in these diagrams. Since our field of view is larger,
the field star contamination is more important and precludes a clear identification of all the CMD features. We can recognize a prominent
MS that  extends from V$\sim$ 16.5 down to 20.5. This last limit does not depend on completeness or photometric depth
but is only set by the constraint to have plotted all the stars having UBVI measures.\\
While the RGB and the clump  (at about V$\sim$14.5) are well delineated,
the subgiant branch is clearly blurred
and confused by field stars. 
The field star MS crosses the subgiant branch and continues up to V$\sim$13.0.\\
The blue straggler stars sequence is also well defined at V$\sim$16.0, (B-V) $\sim \leq$ 0.3 .
Finally, on the right side of the MS, a conspicuous binary sequence is present,
as previously outlined by Zlocewski et al. (2007).\\

We make use of the results in the previous section to alleviate the field-star contamination and to have a better handle on the key features
of the CMD:  the TO, the subgiant branch, the RBG and the clump.
With this aim, we select all the stars within the cluster radius, and derive the CMD shown in Fig.~6 in the B/U-I plane.
We use this color combination to have  a better view of both blue and red parts of the CMD.
The MS is very clean and extends down to B$\sim$21.0. Close to the TO, between B$\sim$17.5 and B$\sim$ 18, the MS broadens,
but this broadening is mostly produced by the intersection with the binary sequence, which is well clean down to  B$\sim$21.
The cluster TO is then located at B$\sim$18, U-I $\sim$1.4 . The subgiant branch and the RGB are more scattered, but the bottom of the RGB is
most probably located at B$\sim$17.7, U-I$\sim$2.65. The RGB clump is spread in color, and tilted along the reddening vector, which implies that
some differential reddening must be present (Carraro et al. 2002).  \\

\noindent
To quantify the effect, photometry has been corrected  by means of a procedure described in full detail in Milone et al. (2012). 
Briefly, we iteratively define a fiducial MS for the cluster and then compute the displacement along the reddening vector of each star from this
fiducial line.
This systematic color and magnitude offset are our estimates of the local differential-reddening values. A map with the resulting reddening variation
is shown in Fig~7. This map indicats that differential reddening is present, and its maximum spatial variation amount to 0.07 mag,  which  is significantly lower
than Zlocewski et al. (2007) estimate, that was based, however, on the low resolution  Far InfraRed Background(FIRB)  maps (Schlegel et al.  1998).  In Fig.~8 we show a zoom
of the MS region in Melotte~66 CMD before and after the variable reddening correction. The corrected MS is clearly less wide than the original, and
the overall quality of the CMD improves significantly.
Based on these results,  the MS broadening can be  entirely accounted for by the presence of a certain amount of binary stars and variable extinction across the cluster field.
Other scenarios, like extended star formation or metallicity spread, are therefore not required  to explain the MS natural width.\\

  \begin{figure}
   \centering
   \includegraphics[width=\columnwidth]{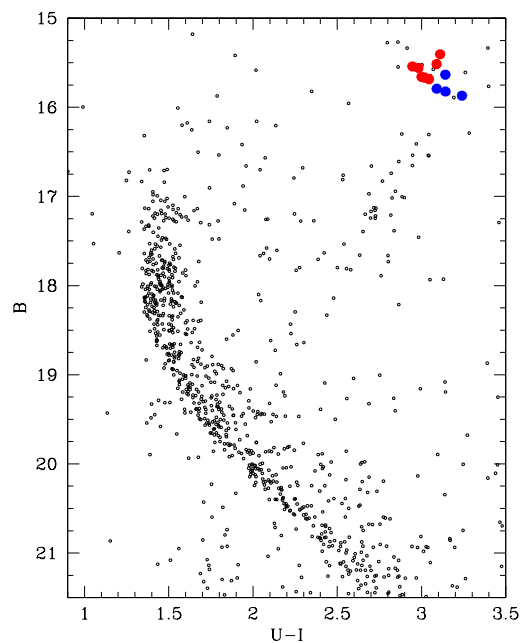}
   \caption{CMD of Melotte~66. Only stars within 6 arcmin are shown. Clump stars are color-coded: we could obtain high-resolution spectra for the red stars, while the blue
   are taken from Sestito et al. (2008). }
    \end{figure}

  \begin{figure}
   \centering
   \includegraphics[width=\columnwidth]{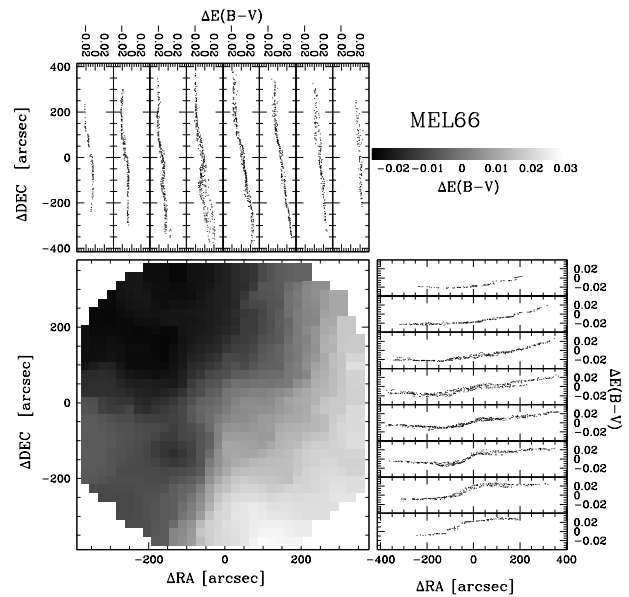}
   \caption{A reddening map in the region of Melotte~66. See text for details}
    \end{figure}

 \begin{figure}
   \centering
   \includegraphics[width=\columnwidth]{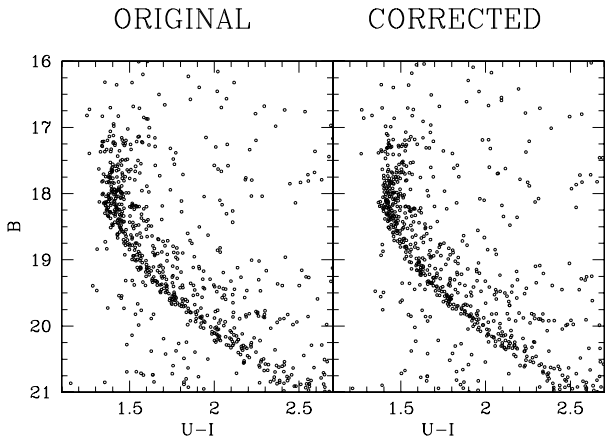}
   \caption{A zoom of the MS region for Melotte~66 before (left panel)  and after differential reddening correction (right panel).}
    \end{figure}

  \begin{figure}
   \centering
   \includegraphics[width=\columnwidth]{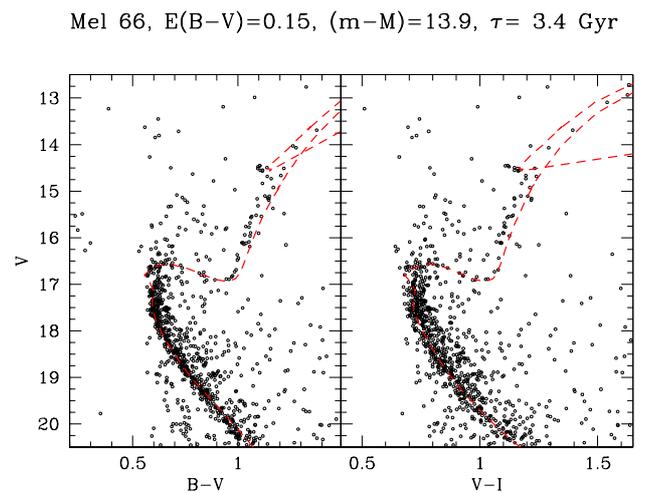}
   \caption{Isochrone solution for Melotte~66 in the V/B-V (left panel) and V/V-I (right panel) planes. The best fit has been obtained
   for an age of 3.4 Gyr, E(B-V)=0.15, and (m-M)=13.9 . See text for additional details.}
    \end{figure}

\noindent
To confirm this scenario further, we make use of synthetic CMDs, as generated from the Padova (Bressan et al. 2012) suite of models. The method is described in detail
in Carraro et al. (2002). First of all, we need an estimate of the cluster fundamental parameter. With this aim, we make use of isochrone for the metallicity Z =0.009, which is 
derived from the [Fe/H] valued obtained in this paper (see below). 
This is illustrated in Fig.~9, in both the V/B-V and V/V-I plane. The fit is generally good in the MS and TO region. The magnitude of the clump is also well reproduced. There is, however, a clear problem with the color of the RGB, that models predict redder than observations.  The best-fit isochrone is for an age of 3.4$\pm$0.2 Gyr.  The uncertainty in the age has been derived by super imposing
many different isochrones and evaluating by eye whether they produced a reasonable fit or not.\\
For this estimate of the age, we infer a reddening of E(B-V)=0.15$\pm$0.03, which is in line with previous estimates, and an apparent distance modulus ($m-M$)= 13.9$\pm$0.2. From these two figures we derive an absolute distance modulus ($m-M_0$)=13.4$\pm$0.3.
In turn, this is used to estimate an heliocentric distance of 4.7$^{+0.2}_{-0.1}$ kpc for Melotte~66.\\
Therefore, Melotte~66 is one of the old open clusters with the largest displacement  ($\sim$ 1.0kpc) from  the formal Galactic plane. 
This, however,  has to be considered as an upper limit, since the disk is significantly warped in the cluster  direction (Moitinho et al. 2006).

\noindent
With this information at hand, we generated synthetic CMDs by using the {\it TRILEGAL} code (Girardi et al. 2005) and by 
following Carraro et al. (2002, 2006) prescriptions. First of all we generate a synthetic star cluster for the age,
reddening, and distance of Melotte~66, by adding 30\% of binaries with mass ratio in the range 0.6 to 1.0. This is shown in Fig.~10, upper left panel (a). 
In the upper right panel (b) we show a realization of the expected Milky Way population in the direction of Melotte~66  for an area as large as the one
covered by our photometry. The simulation includes stars from the Galactic halo (color-coded in red),  and the Galactic thin and thick disk (color-coded in
green and blue, respectively). No stars from the Galactic bulge are expected for this specific direction.\\
In the middle panels ((c) and (d)), we show the same two CMDs blurred by the errors as  in our photometric data set. Finally, the lower left panel (e) illustrates
the combination of the synthetic cluster plus the synthetic Galactic field, which one has to compare with the real data in the lower right (f) panel.\\
Overall, the two last panels look very similar, which means that we modeled correctly the field star population
statistically speaking, and our ingredients,
which may  distance, reddening, metallicity and binary properties, are mostly fine. These simulations show that  there are many thick disk stars  in the line of sight to Melotte~66   .
They form a sequence which intersect the cluster sub giant branch. Halo and thin disk stars are a minor contribution. 
It is also evident that field stars do not  have colors bluer than the cluster TO, and therefore,  there are no field stars in the region of the CMD where
blue stragglers are located.

  \begin{figure}
   \centering
  \includegraphics[width=\columnwidth]{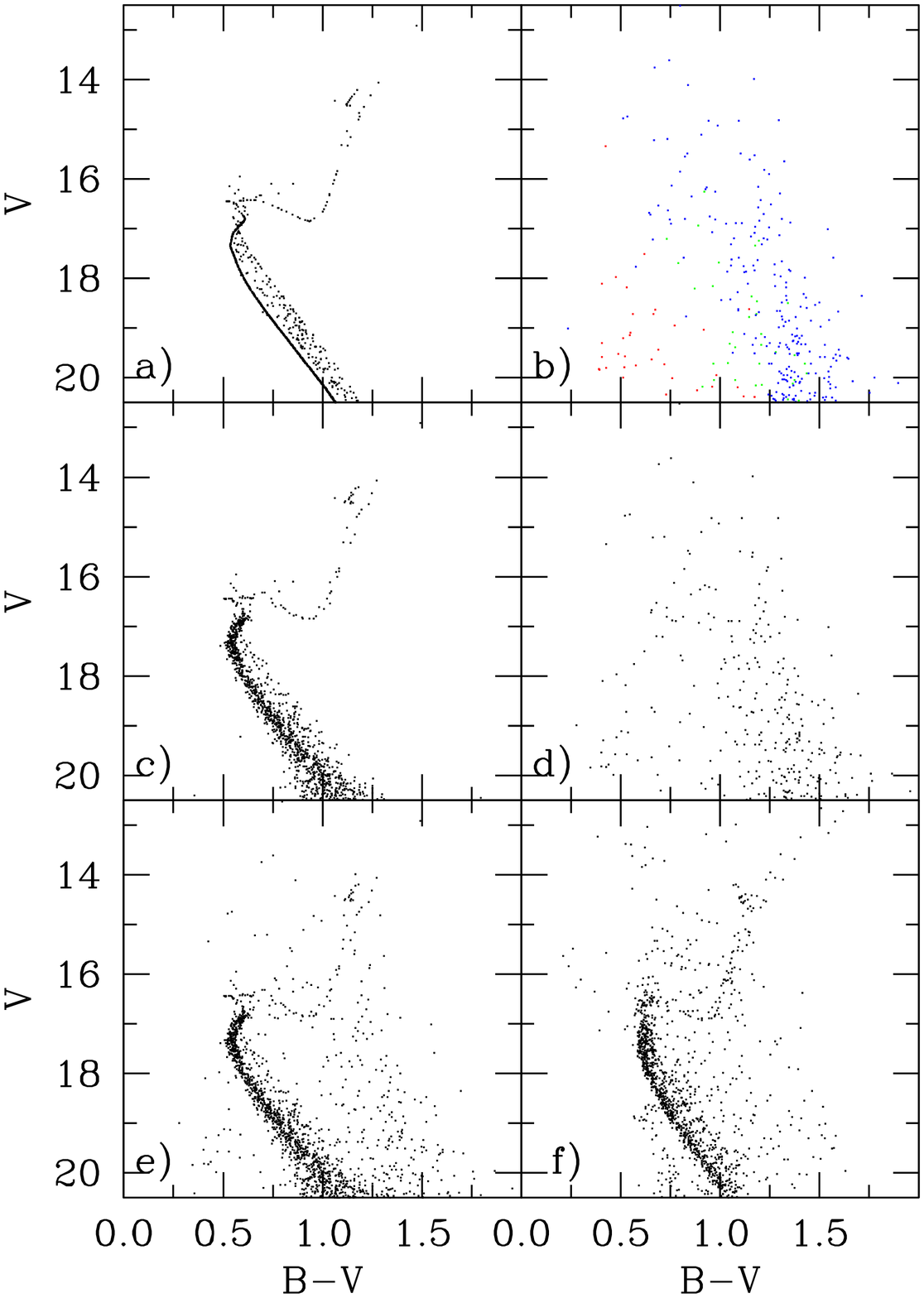}
   \caption{A simulation of the stellar field in the line of sight toward Melotte~66. See text for more details.}
    \end{figure}

  \begin{figure}
  \centering
   \includegraphics[width=\columnwidth]{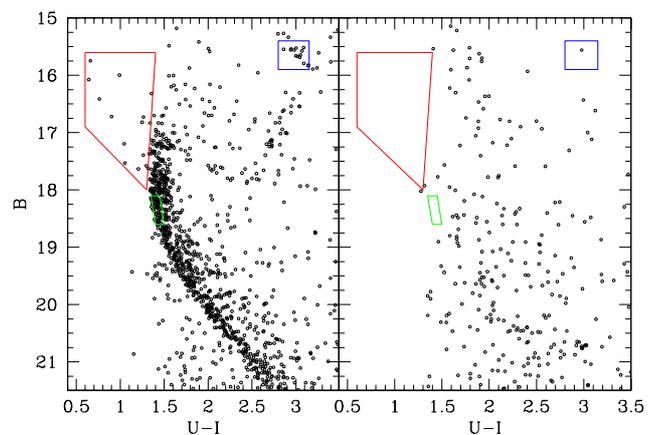}
   \caption{Selection of BSS, MS,  and red-clump stars in Melotte 66 CMD (left panel). An equal area field is shown in the right panel to probe field star contamination}
    \end{figure} 

  \begin{figure}
  \centering
   \includegraphics[width=\columnwidth]{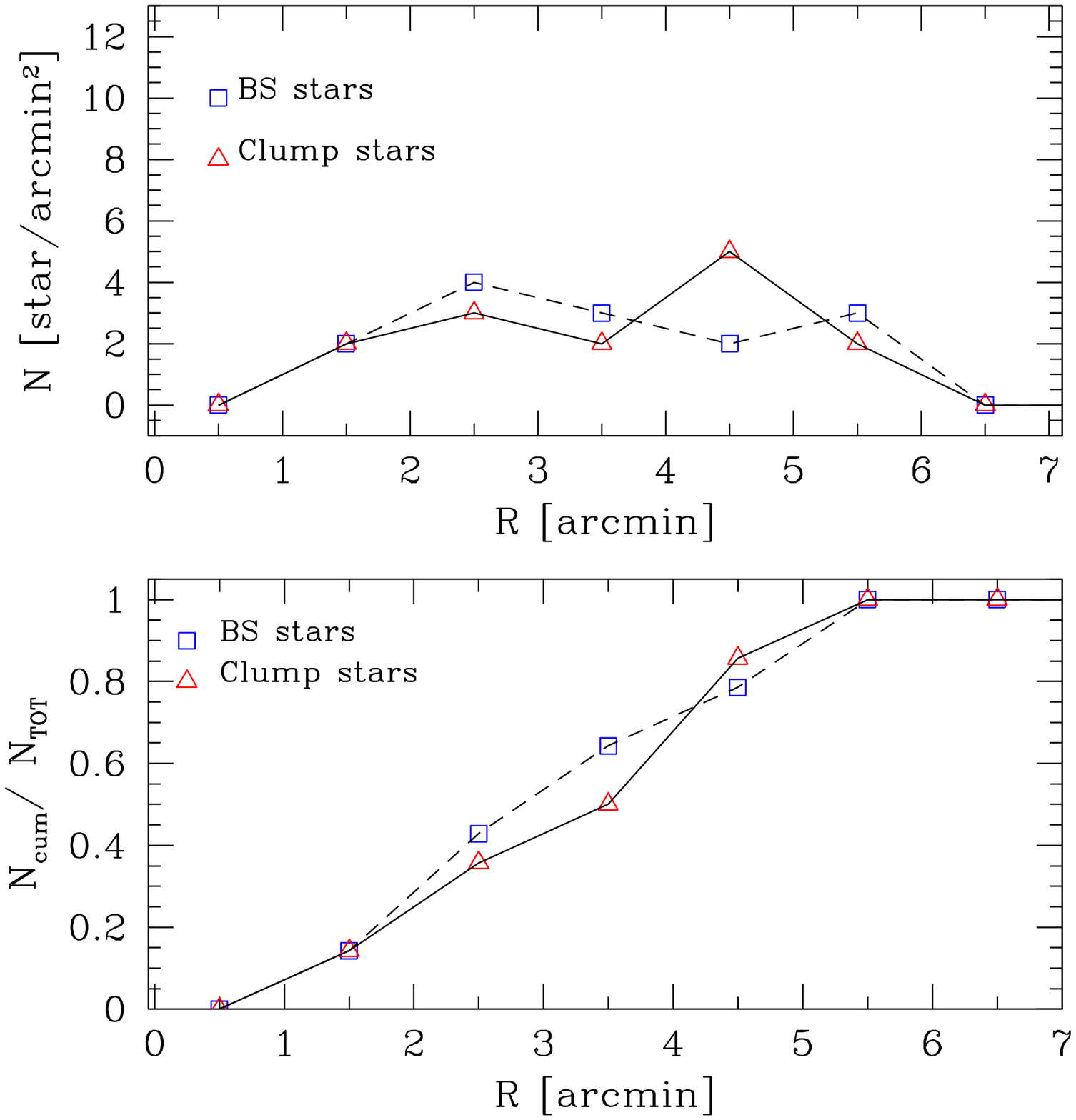}
   \caption{Upper panel: Radial distribution of BSS and clump stars. Lower panel: Cumulative radial distribution of BSS and clump stars.}
    \end{figure} 
    
 \begin{figure}
 \centering
   \includegraphics[width=\columnwidth]{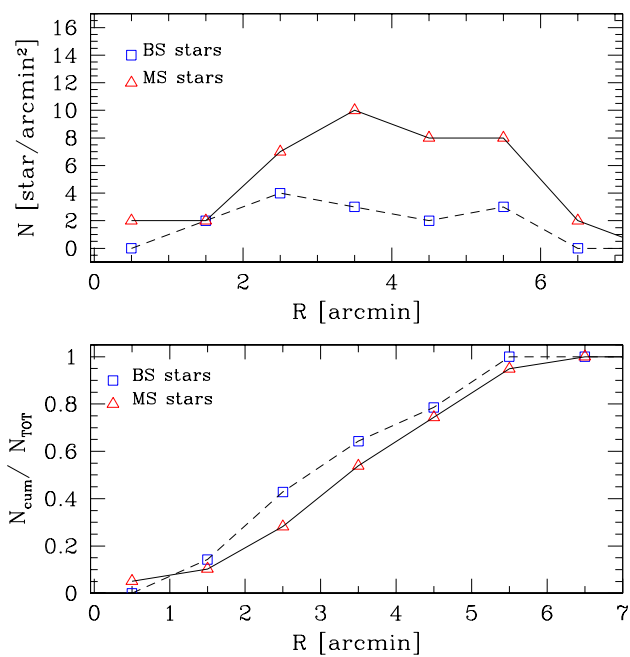}
   \caption{Upper panel: Radial distribution of BSS and MS stars. Lower panel: Cumulative radial distribution of BSS and MS stars.}
    \end{figure}

\section{Blue straggler stars in Melotte 66}
As mentioned in Section~3, it has been  known since the early study from Hawarden (1976) that Melotte~66 harbors a population of blue straggler stars (BSS).  According to the most recent
compilation by Ahumada \& Lapasset (2007),  the cluster hosts as many as 35 BSS, while De Marchi et al. (2006)  found 29 BSS within a radius of 7 arcmin from the cluster center. 
However, as emphasized by Carraro et al. (2008), the precise number of
BSS in open clusters is difficult to know because of the severe field star contamination. As a consequence of this, many field stars fall in the region of the CMD,
where BSS are routinely found, which  significantly affects their statistics. In the case of Melotte~66, field star contamination is not as important  as in many other open clusters because
of its relative high Galactic latitude. Still, as shown in Fig~8, some field star contamination is present.  This is in the  form of a tilted sequence, which crosses the subgiant branch of Melotte~66,
and it is mostly composed by thick disk stars (see Sect.~5)\\

\noindent
We start the analysis of the BSS population by selecting BSS candidates in the CMD  with a sample of clump and MS stars. This is illustrated in Fig.~13, where the left panel shows the BBS candidate region (red polygon), the clump region (blue square), and a sample of MS single stars (green polygon). In the right panel, an equal area field realization is shown,
to see that the selected stars fall in regions that are not significantly contaminated by field stars. MS stars, in particular, have been searched for in a region of the MS not affected by
incompleteness and un a region that is most probably free of binary star contamination. As a result, we find 14 candidate BSS and compare them with a sample of {\it bona fide} clump  stars (14)
and MS single stars (39).\\
The comparison with clump stars is shown in Fig~14, while the comparison with MS stars is shown in Fig~15. In both figures, the upper panels present the radial density profile of the two
populations, while the lower panels present the radial profile of the normalized cumulative distribution (see Carraro \& Seleznev 2011 for more details). Clump stars and BSS 
do show about the same distribution, and indeed a Kolmogorov-Smirnov (KS) test  implies a probability of 79$\%$ when they are drawn from the same parent distribution. There seems to be
marginal evidence that BSS are more concentrated than clump stars.
A completely different scenario is revealed by  comparing  MS stars (see Fig.~14). In this case, it seems clear that  the BSS are more centrally concentrated than MS single stars,
and the KS test in this case returns a 20$\%$ probability that they have the same origin. 
This analysis lends support to a scenario where BSS stars in Melotte~66 are most probably binaries. The environment of Melotte~66 is quite loose, and therefore they might be primordial binaries that sank toward the cluster center because of their larger combined mass, and then survived in the cluster central regions. A spectroscopic study of these stars would be very useful to support or deny our conclusions.

\section{Observation and data reduction : Spectroscopy}
Observations were taken in service mode on the nights of February 12 and March
4, 2012 using the multi-object, fiber-fed FLAMES facility mounted at the
ESO-VLT/UT2 telescope at the Paranal observatory (Chile).  Two 2400s exposures
were taken with the red arm of the UVES high-resolution spectrograph
setup at 5800\AA~, a central wavelength that covers the 4760-6840\AA\, wavelength
range and  thus provides a resolution of R$\simeq$47000. 

The data were reduced using the ESO CPL based FLAMES-UVES pipeline version
5.0.9\footnote{\url{http://www.eso.org/sci/software/pipelines/}} for
extracting the individual fiber spectra. 

The spectra were eventually normalized using the standard IRAF task {\tt
continuum}. Radial velocities were computed using the IRAF/{\tt fxcor} task to
cross-correlate the observed spectra with a synthetic one from the
\citet[][]{coelho05} library with stellar parameters  $T_{\rm eff}$=5250\,K, log\,$g$=2.5, solar
metallicity, and no $\alpha$-enhancement. The IRAF {\tt rvcorrect} task was used
to calculate the correction from geocentric velocities to heliocentric. We took  the star's radial
velocity to be the average of the two epochs measured and the error to be the difference
between  the two values multiplied by 0.63 (see Keeping 1962).

Finally, for the abundance analysis, the two epoch rest-frame spectra obtained for each star were averaged together.
The final spectra have signal-to-noise (SNR) ratios in the range 30-50 at $\sim$6070\AA.

\section{High resolution chemical abundances}

\subsection{Abundance analysis}

Our spectroscopic sample consists of seven targets from UVES observations and six targets from Sestito et al. (2008), with six  stars in common. The elemental abundances were derived based on
equivalent width (EW) measurements using the MOOG abundance calculation code (Sneden 1973). The EWs were measured using the automated ARES code (Sousa et al. 2007) with frequent manual
checks of the EWs. Interpolated Kurucz model atmospheres based on the ATLAS9 code (Kurucz 1993, Castelli 1997) with no convective overshooting were used throughout the analysis. \\

We derive the stellar parameters ( T$_{\rm eff}$, log$g$, and $\xi$ ) based on spectroscopy. Abundances for all Fe {\sc i} and {\sc ii} lines were computed from the measured EWs, where we  always adopts a starting model of  T$_{\rm eff}$= 4850 K and log$g$, = 2.5. 
The effective temperature was derived by requiring excitation equilibrium of the Fe~{\sc i} lines. Micro-turbulence was derived from the condition that abundances from Fe {\sc i} lines show no trend with EWs. Surface gravity was derived via ionization equilibrium, which requires the abundances from Fe {\sc i} lines to equal those from Fe {\sc ii} lines. The adopted stellar parameters are shown in Table~\ref{t:params} with the measured radial velocity.\\

Next, we derived elemental abundances for Na,  Mg, Al, Si, Ca, Ti, Cr, Ni, and Ba by adopting the solar abundance values from 
Grevesse and Sauval (1998) when calculating the relative abundances. Final abundances are reported in Table~\ref{t:results} along with the corresponding line-to-line standard deviation (rms) from the mean abundances. 
These results are shown graphically in Fig.~11. \\

\subsection{Error budget}

Chemical abundances are largely affected by two sources of uncertainties: $(i)$ error in the EW measurement and $(ii)$ error in the stellar parameters (T$_{\rm eff}$, log$g$, and $\xi$). 
There are also uncertainties in the atomic data, known as the  log $gf$, however,m the effect of this is negligible when looking for star-to-star variations given the narrow range of stellar parameters in this sample, where any such effect would systematically affect all stars and not contribute to the random errors internal to this study. The random errors in the abundances due to EW measurements are well represented by the standard deviation (rms) from the mean abundance based on the entire set of lines.\\

Abundance errors due to stellar parameters were estimated by varying one parameter at a time and checking the corresponding variation in the resulting abundance. We adopted variations of $\pm$50~K in T$_{\rm eff}$ and $\pm$0.05 km$^{-1}$ in $\xi$ because larger changes in those quantities would have introduced a significant trend in $\log~n$(Fe) $vs$ the excitation potentials and the line strength, respectively. The uncertainties in log($g$) were estimated by varying this quantity until the difference between $log~n$(Fe{\sc i}) and $log~n$(Fe{\sc ii}) is larger than 0.1~dex, when the ionization equilibrium condition is no longer satisfied. The typical error in log($g$) was 0.1 dex. The effect of stellar abundances due to stellar parameters are presented in Table \ref{t:errors}. Over all,  the total error is dominated by the line-to-line scatter with minimal impact from the error in stellar parameters.\\

\begin{table*}
\caption[]{Stellar parameters for all the spectroscopic targets. IDs in column 1 are from this study.}
\begin{tiny}
\begin{tabular}{lccccccccc}
\hline\hline
ID & Kassis et al. & Sestito et al. & T$_{\rm eff}$ &  logg & $\xi$  &  [Fe/H]  & RV \\
       &                      &                        &    (K)                     &     & kms$^{-1}$    &     & kms$^{-1}$ \\
\hline
597    & 385	&  ---      & 4900 & 1.80	& 1.45	& -0.30$\pm$ 0.07    & 22.63$\pm$0.03 \\
776    & 603	&  ---      & 4800	 & 2.00	& 1.63	& -0.27$\pm$ 0.09  & 21.26$\pm$0.07 \\
1521  & 1419	& 2218  & 4900	 & 2.00	& 1.35	& -0.30$\pm$ 0.08  & 21.29$\pm$ 0.14 \\
2099  & 1953	&1346   & 4850	 & 1.90	& 1.37	&- 0.28$\pm$ 0.09  &  22.58$\pm$ 0.65\\
2209  & 2155	&  ---      & 4850	 & 2.00	& 1.45	& -0.26$\pm$ 0.08  & 22.52$\pm$0.03\\ 
2291  & 2187	&  ---      & 4850	 & 1.90	& 1.46	& -0.26$\pm$ 0.09  & 20.13$\pm$0.03\\  
2803 & 2771	& 1785  & 4850	 & 2.00	& 1.38	& -0.26$\pm$ 0.09  & 22.54$\pm$ 0.14 \\ 
1202 & 1000   &1493    & 4850	& 2.00	& 1.30	& -0.30$\pm$ 0.09   & 22.90$\pm$ 0.05\\
2980 & 2945   &1865    & 4850	& 2.00	& 1.45	& -0.32$\pm$ 0.09  & 18.76$\pm$ 0.03 \\
1919 & 1805  &1884    &  4850	& 2.50	& 1.30	& -0.22$\pm$ 0.09 &  21.92$\pm$ 0.03\\
\hline
\hline
\end{tabular}
\end{tiny}
\label{t:params}
\end{table*}

\begin{table*}
\caption[]{Mean abundances and standard deviations.}
\begin{tiny}
\begin{tabular}{cccccccccccccc}
\hline\hline
Kassis et al. & Sestito et al.    &[Na/Fe]     &     [Mg/Fe]    &     [Al/Fe]     &   [Si/Fe]     &     [Ca/Fe]     &      [Ti/Fe]  &     [Cr/Fe]  &  [Ni/Fe]  &        [Ba~{\sc ii}/Fe]  \\
\hline
385	  &  ---      &  0.07 $\pm$0.07    &  0.19$\pm$0.01    &     0.18 $\pm$0.07   &     0.05 $\pm$0.08  &    0.10$\pm$0.09    &   -0.05 $\pm$0.07   &    -0.05 $\pm$0.08    &   -0.08$\pm$0.08    & 0.35 $\pm$0.03\\
603	  &  ---      & 0.03$\pm$0.03     &  0.14$\pm$0.05    &     0.16$\pm$0.03    &     0.01$\pm$0.07   &    0.03$\pm$0.07     &   -0.08$\pm$0.11    &   -0.13$\pm$0.06      &    -0.07$\pm$0.08   &  0.29$\pm$0.01\\
1419	  &  2218 & 0.06$\pm$0.06     &  0.22$\pm$0.02    &     0.17 $\pm$0.07   &     0.02$\pm$0.06   &    0.08$\pm$0.09     &   -0.09$\pm$0.10    &    -0.07$\pm$0.06     &    -0.07$\pm$0.08   &   0.33$\pm$0.01\\
1953	  &  1346 & 0.07$\pm$0.03     &  0.16$\pm$0.06    &     0.16$\pm$ 0.08   &     0.11$\pm$0.06   &   0.03$\pm$0.04      &   -0.10$\pm$0.09    &    -0.05$\pm$0.07     &    -0.09$\pm$0.09   &  0.32$\pm$0.01\\
2155	  &   ---      & 0.03$\pm$0.03     &  0.17$\pm$0.04    &     0.10$\pm$0.03      &     0.08$\pm$0.08   &   0.02$\pm$0.08      &    -0.07$\pm$0.07   &     -0.09$\pm$0.07    &     -0.08$\pm$0.08  &   0.30$\pm$0.08\\
2187	  &   ---      & 0.05$\pm$0.06     &  0.13$\pm$0.01    &     0.18$\pm$0.04    &     0.03$\pm$0.10     &   0.05$\pm$0.10        &    -0.08$\pm$0.08   &    -0.07$\pm$0.05     &    -0.04$\pm$0.06   &   0.30$\pm$0.02 \\
2771	  &  1785 & 0.04$\pm$0.03     &  0.15$\pm$0.01    &     0.16$\pm$0.06   &     0.08$\pm$0.08   &   0.04$\pm$0.07    &     -0.06$\pm$0.07   &   -0.09$\pm$0.10        &    -0.03$\pm$0.08    &  0.35$\pm$0.02 \\
 1000 & 1493  &  0.09$\pm$0.08     &  0.18$\pm$0.02    &     0.16$\pm$0.04    &     0.05$\pm$0.09   &   0.07$\pm$0.10        &    -0.10$\pm$0.08     &      -0.10$\pm$0.07     &     -0.08$\pm$0.08   &  0.37$\pm$0.02 \\
 2945  & 1865  &  0.32$\pm$0.06     &  0.21$\pm$0.08    &     0.18$\pm$0.10      &     0.09$\pm$0.10     &   0.03$\pm$0.09      &   -0.10$\pm$0.08      &    -0.04$\pm$0.08     &   -0.07$\pm$0.08    &   0.52$\pm$0.03 \\
  1805  & 1884  &  0.01$\pm$0.05     &  0.10$\pm$0.03   &     0.07$\pm$0.09    &   -0.02$\pm$0.10      &   0.01$\pm$0.10        &   -0.08$\pm$0.09    &   -0.04$\pm$0.06     &  -0.07$\pm$0.09     &   0.26$\pm$0.06 \\
\hline
\hline
\end{tabular}
\end{tiny}
\label{t:results}
\end{table*}

\begin{table*}
\caption{Errors due to stellar parameters.}
\begin{tiny}
\begin{tabular}{lccccccccccccc}
\hline
Parameter & Fe & Na     &   Mg   &     Al  &   Si  &   Ca    &    Ti &   Cr  &  Ni  &   Ba  \\
\hline 
$\Delta$ $T_{\rm eff}$ $\pm$ 50K &  $\pm$0.03&  $\pm$0.04   &   $\mp$0.03    &   $\pm$0.03      &   $\mp$0.01        &   $\pm$0.05    &   $\pm$0.07     &  $\pm$0.06     &   $\pm$0.02 & $\pm$0.01 \\
$\Delta$log$g$ $\pm$ 0.1              & 0.00            & 0.00  &   0.00     &  0.00      &   $\pm$0.02        &   $\mp$0.01   &   0.00   &  0.00   &   $\pm$0.01 & $\pm$0.03 \\
$\Delta$ $\xi$ $\pm$ 0.05               & $\mp$0.03&  $\mp$0.01   &   $\mp$0.01    &   0.00      &   $\mp$0.01        &   $\mp$0.03    &   $\mp$0.02     &  $\mp$0.02     &   $\mp$0.03 & $\mp$0.04 \\

\hline
\end{tabular}
\end{tiny}
\label{t:errors}
\end{table*}

\subsection{Literature comparison}    

We compare our derived parameters and abundances against those derived by Sestito et al. (2008). We share six stars in common, and overall, the results are consistent within the quoted error budget. A detailed look at the differences are presented in Table~\ref{t:diff}. For elements heaver than Si only, the quoted average abundances are compared; as for Na, Mg and Al, we compared the individual lines common to both studies  and the corresponding LTE abundances.\\

For stellar parameters, our effective temperature values are hotter by no more than 100K, which is not unreasonable given the various differences in the temperature calculations. For most
stars, surface gravity also agrees within expected uncertainties with, perhaps, the exception of star 2218, for which our value is significantly lower than in Sestito et al. (2008). Other notable difference is the Al abundance, where  our values are larger than expected within the quantified uncertainties  for stars 1346, 1493 and 1785  . It is unclear what the source of this discrepancy is, and the most likely case could be the continuum normalization of the spectra and its effect on the measured EW.  Sestito et al (2008) noted these large line-to-line variations for Na, Mg, and Al; hence we made the decision to compare abundances per individual line. Despite the different approaches, both studies reach the same conclusion that there are no significant star-to-star abundance variations within the sample.

\subsection{Discussion of abundances} 

The abundance analysis of ten red clump stars in Melotte 66 indicates that this sample is very much chemically homogeneous with little abundance scatter within the cluster members. 
One star (1865) is found to deviate from the cluster mean values for Na
and Ba. This star has a radial velocity outside of the
cluster velocity could potentially be a non member. It is also
possible that this is a cluster binary star, where mass transfer is
triggering an over production of Na and Ba (Sneden et al. 2003).
Excluding this star, the mean abundances and the standard deviation of nine
clump stars in Melotte 66 are given in Table~\ref{t:mean}. \\

The abundance patterns of heavy elements observed from stellar photospheres is likely to be a signature of its natal proto-cluster cloud composition with  intrinsic star-to-star variations within a cluster that are expected to be close to zero (De Silva et al. 2006). Our results show that Melotte 66 has little or no intrinsic star-to-star abundance variations, supporting the scenario that it is a typical open cluster in the Galactic disk that is born out of a uniformly mixed proto-cluster gas cloud. \\

 \begin{figure}
 \centering
\includegraphics[width=\columnwidth]{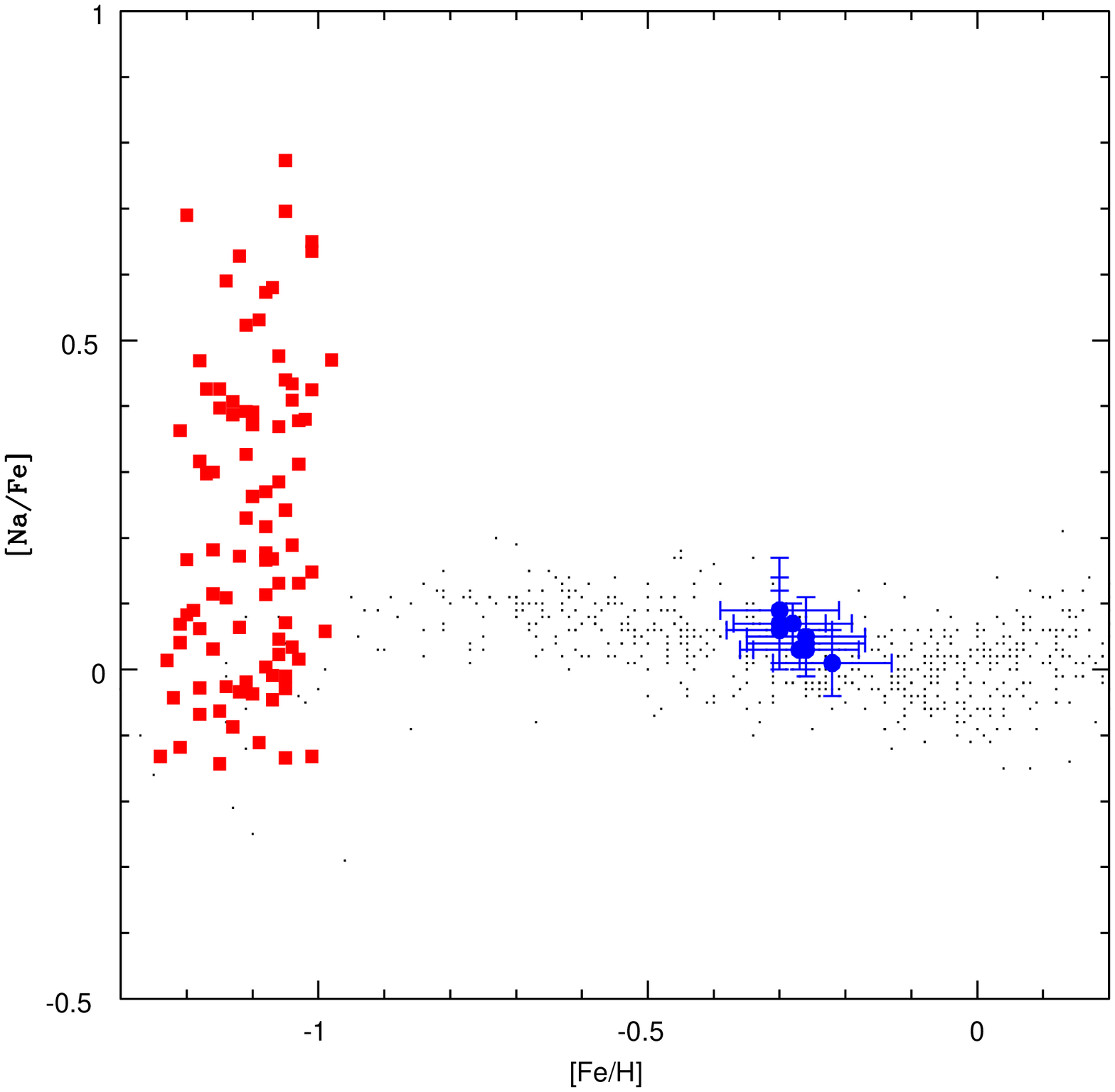}
 \caption{[Na/Fe] vs [Fe/H]. The large filled circles are Melotte 66 sample stars. The red squares are NGC 2808 stars from Carretta et al. (2006), and the black points are thin and thick disk stars from Bensby et al. ( 2014).}
 \label{nafe}
 \end{figure}

 \begin{figure}
 \centering
\includegraphics[width=\columnwidth]{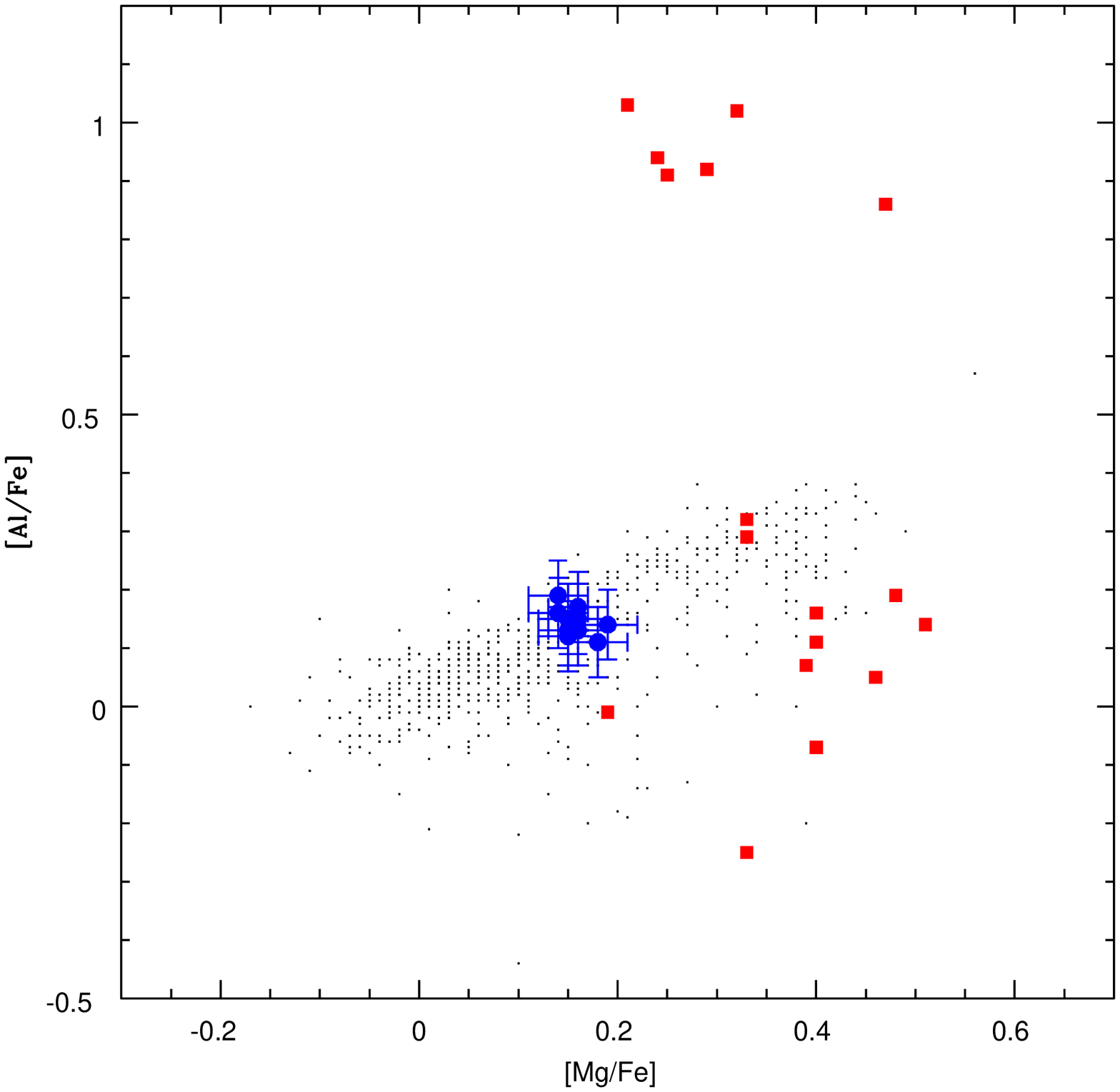}
 \caption{[Al/Fe] vs [Mg/Fe]. The large filled circles are Melotte 66 sample stars. The red squares are NGC 2808 stars from Carretta et al. (2006) and the black points are thin and thick disk stars from Bensby et al. ( 2014).}
 \label{mgal}
 \end{figure}

To further explore the apparent lack of multiple stellar populations, we plot [Na/Fe] vs. [Fe/H] (Figure \ref{nafe}) and [Mg/Fe] vs. [Al/Fe] (Figure \ref{mgal})  with the abundances of template globular cluster NGC 2808 (red squares  from Carretta et al. 2006) and disk field stars (black points, from Bensby et al. 2014). The plots shows very clearly that Melotte 66 abundances do not reach the extreme chemical enhancements that indicate multiple stellar populations in globular clusters.\\

We now compare the abundances of Melotte 66 with other clusters studied with high resolution spectroscopy. Figure~\ref{ocs} shows the average Melotte 66 abundances from this study against other open cluster average abundances from Pancino et al. (2010).  The cluster sits within the general spread of the open cluster ratios, and no unusual variations are seen.   \\

\begin{figure}
 \centering
\includegraphics[width=\columnwidth]{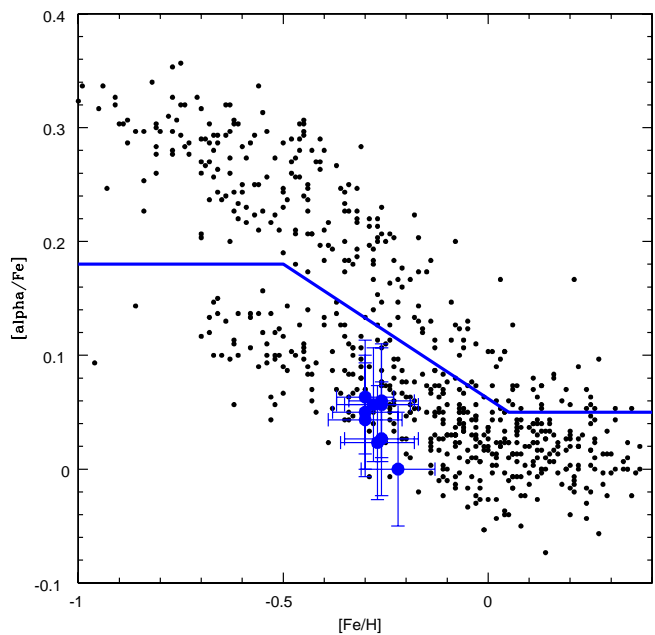}
 \caption{[$\alpha$/Fe] vs [Fe/H] for Melotte 66 sample stars (blue circles). The black points are thin and thick disk star from Bensby et al. ( 2014), and the line dividing the thin and thick disk is from Haywood et al. 2013.}
 \label{disks}
 \end{figure}

 We now explore the cluster abundances against the characteristic abundances of the thin and thick disk to better understand the cluster origin. In Figure \ref{disks}, we plot the mean [$\alpha$/Fe] against [Fe/H] for the cluster and Bensby et al. (2014), where $\alpha$ elements represent the average abundances of Mg, Si, and Ti. The plot also shows the suggested thin-thick disk division line by Haywood et al. (2013) to better guide the eye. It is clear that Melotte 66 sits within the general thin disk population rather than the thick disk.\\

Our abundance results show that the alpha elements Si, Ca, and Na are close to solar ratios in Melotte 66, while Mg and Al are enhanced by $~$0.15 dex relative to Fe. It is also clear that Ba, the only neutron capture element, is enhanced in this cluster. All these elements are, however, within the trend described {\bf by disk field stars.} \\

\begin{figure}
 \centering
\includegraphics[width=\columnwidth]{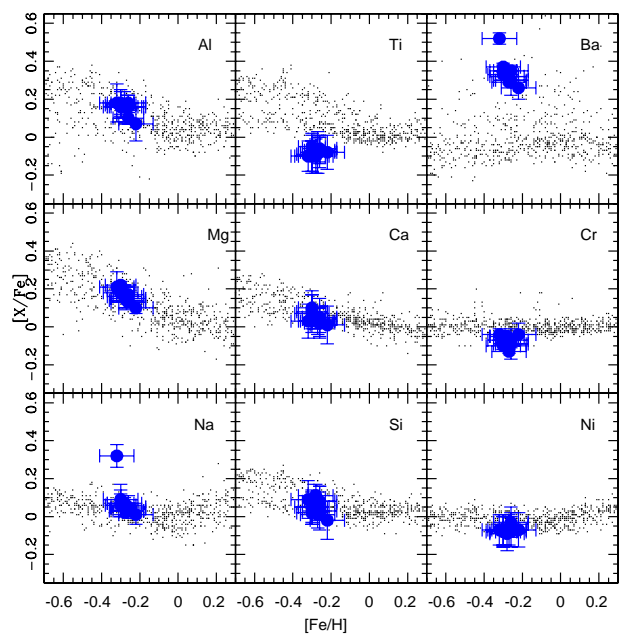}
 \caption{Abundance ratios in the form of [X/Fe] vs [Fe/H]. The large filled circles are Melotte 66 sample stars. The black points are thin and thick disk stars from Bensby et al. ( 2014).}
 \label{grid9}
 \end{figure}

\begin{figure}
\centering
\includegraphics[width=\columnwidth]{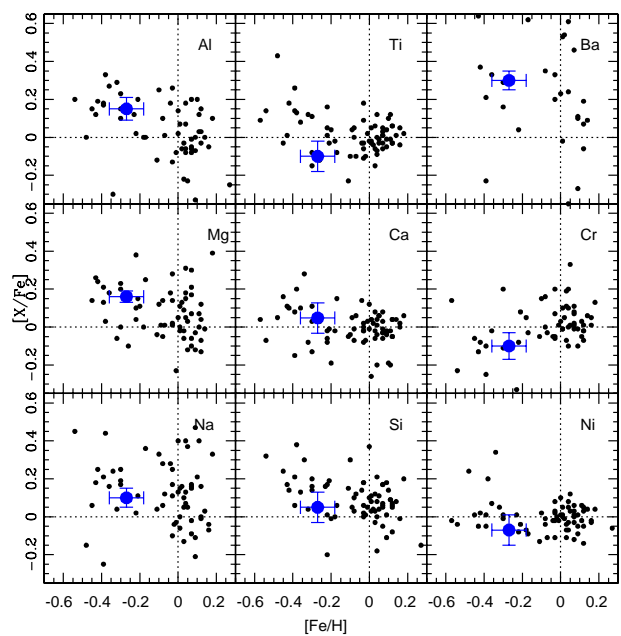}
 \caption{Average abundance ratios of Melotte 66 (blue circle) compared to literature cluster values from Pancino et al. (2010). The cross-hairs highlight the solar abundances.}
 \label{ocs}
 \end{figure}

The remaining elements, Ti, Cr, and Ni are consistently under abundant relative to Fe by about $\sim$0.07 dex. This was also seen in the Sestito et al (2008) study, suggesting that this is a feature of the chemical abundance pattern of Melotte 66. In Figure~\ref{grid9}, it is clear that the cluster stars sit below the bulk of the thin  disk stars for Ti and also marginally for Cr and Ni. Similar abundance patterns where Ti behaves more like an Fe-peak element, rather than an alpha-capture element have been observed among other clusters as well (e.g. NGC 2324 and NGC 2477 (Bragaglia et al 2008)). Ti remains an intriguing element with its nucleosynthesis origins that are not fully understood.\\

While a given open cluster is highly chemically homogeneous internally, this highlights abundance variations are seen among the open cluster population (see also De Silva et al 2009). The likely explanation for such different abundance patterns across the open clusters is due to the different star-formation and chemical feedback histories for different regions of the Galaxy.  Such differences are key to chemically identifying stars that formed together during the build up of the Galaxy, which are now dissolved into a the general field population (Mitschang et al. 2014). \\

\begin{table} 
\caption{Comparison with Sestito et al 2008 for the common stars. Only the Sestito star ID is given.}
\begin{tabular}{lccccccc}
\hline
 & 1346 & 1493 & 1785 & 1865 & 1884 & 2218 \\
\hline 
$\Delta$ $T_{\rm eff}$ &  100 & 80 & 80 & 133 & 100 & 50 \\
$\Delta$log$g$& 0.1 & 0.15 & 0.05 & 0.05 & 0.05 & 0.39 \\
$\Delta$ $\xi$ & 0.2 & 0.1 & 0.18 & 0.21 & 0.07 & 0.1 \\
$\Delta$ Na: 5688 & 0.06 & 0.08 & 0.00 & -0.02 & 0.19 & -0.01 \\
$\Delta$ Na: 6154 & 0.09 & -0.06 & 0.05 & 0.03 & 0.00 & 0.00 \\
$\Delta$ Na: 6160 & 0.04 & 0.10 & -0.03 & -0.14 & -0.04 & -0.05 \\
$\Delta$ Mg: 6318 & -0.03 & -0.01 & -0.05 & -0.08 & -0.05 & -0.01 \\
$\Delta$ Al: 6696 & 0.30 & 0.31 & 0.07 & 0.07 & 0.16 & 0.09 \\
$\Delta$ Al: 6698 & 0.31 & -0.51 & 0.34 & 0.22 & 0.17 & 0.21 \\
$\Delta$ [Fe/H] & -0.09 & -0.05 & -0.04 & -0.02 & -0.08 & -0.01 \\
$\Delta$ [Si/Fe] & 0.02 & 0.09 & 0.02 & 0.05 & 0.12 & 0.14 \\
 $\Delta$[Ca/Fe] &  0.12 & 0.02 & 0.06 & 0.04 & 0.05 & 0.00 \\
 $\Delta$[Ti/Fe]  & 0.08 & 0.08 & 0.07 & 0.05 & 0.16 & 0.10  \\
$\Delta$ [Cr/Fe]  &  0.04 & 0.12 & 0.03 & -0.01 & 0.02 & 0.10 \\
$\Delta$ [Ni/Fe] & 0.07 & 0.07 & -0.02 & 0.05 & 0.07 & 0.10  \\
$\Delta$ [Ba/Fe]   &  0.00 & -0.07 & -0.06 & -0.05 & 0.00 & 0.03 \\
\hline
\end{tabular}
\label{t:diff}
\end{table}

\begin{table*}
\caption[]{Mean abundances and standard deviation}
\begin{tiny}
\begin{tabular}{cccccccccccccc}
\hline\hline
& [Fe/H]  &[Na/Fe]     &     [Mg/Fe]    &     [Al/Fe]     &   [Si/Fe]     &     [Ca/Fe]     &      [Ti/Fe]  &     [Cr/Fe]  &  [Ni/Fe]  &        [Ba~{\sc ii}/Fe]  \\
        &                       &       &   &                &                    &                 &                          &                     &                     &                   &                    &                 &                          \\
\hline
Mean	  &  -0.27    &  0.05   &  0.16    &     0.15   &     0.05  &    0.05   &   -0.08  &    -0.07    &   -0.07   & 0.32\\
std	           &  0.03    & 0.03     &  0.04    &     0.04    &     0.04   &    0.03  &   0.02    &    0.03      &    0.02   &  0.03\\
\hline
\hline
\end{tabular}
\end{tiny}
\label{t:mean}
\end{table*}

\section{Conclusions}
We have presented in this paper a photometric and spectroscopic study of  Melotte~66, one of the most massive old open clusters
in the Milky Way disk.
The most important result of our investigation is that Melotte~66 does not show any evidence, either photometric or spectroscopic,
of distinct sub populations among its stars.
Our photometry demonstrates beyond any reasonable doubt that the MS width is produced by the presence of a significant population of binary stars. The binary sequence intersects the single star MS close to the TO,  
producing the visual effect that the MS is wide.
 For the first time, using numerical simulations, we quantify the binary fraction, which would be not smaller than 30\%.\\

We discussed the cluster photometric properties and revised its fundamental parameters. The age is found  to be 3.4$\pm$0.2 Gyr, which is 
younger than in previous investigations. 

The new spectroscopic material we add fully supports the conclusions from photometry. While confirming previous determinations
of [Fe/H], we did not detect any significant spread in any of the elements we could analyze. Melotte~66 looks like a genuine member
of the old, thin disc population when compared with other disc open clusters and disc field stars.

We finally perform a photometric study of the BSS population in the cluster. We found 14 BBS candidates, a value that is fewer than that found in previous studies, and which we suggest to be primordial binaries.

In conclusion, Melotte~66, like Berkeley~39 (Bragaglia et al. 2012), is a single population star cluster. Although limited by the small number statistics, NGC~6791 seems to be the only open cluster with evidence of multiple stellar populations. The reader, however, has to be warned that NGC~6791 is questioned as a disc star cluster (Carraro 2013; Carrera 2012), since its properties are closer to the bulge
stellar population.

Besides Melotte~66 and Berkeley~39, there are not many  massive old open cluster candidates . With caution in the introduction to this paper we had described  that masses for these clusters are extremely  uncertain, and  one can possibly look at Collinder~261 or Trumpler~5 as possible targets for further studies in this direction.

\begin{acknowledgements}
G. Carraro acknowledges financial support from the Anglo Australian Observatory Distinguished Visitor Program
and from the ESO Director General Discretionary Funds, that allowed a visit to AAO In February 2014, where most of this work was done.    
We warmly thank Russell Cannon and Sarah Martell for useful comments.
\end{acknowledgements}



\begin{thebibliography}{}
\bibitem[Ahumada and Lapasset (2007))]{ah07} Ahumada, J.A,, Lapasset, E., 2007, A\&A, 463, 789
\bibitem[Anthony-Twarog et al. (1979)]{at79} Anthony-Twarog, B.J., Twarog, B.A., McClure, R.D., 1979, ApJ, 233, 188
\bibitem[Anthony-Twarog et al. (1994)]{1t94} Anthony-Twarog, B.J., Twarog, B.A., Sheeran, M., 1994, PASP, 106, 499
\bibitem[Bensby et al. (2014)]{ben14} Bensby, T., Feltzing, S., Oey, M.S., 2014, A\&A, 562, 71
\bibitem[Bragaglia et al. (2008)]{bra08} Bragaglia, A., Sestito, P., Villanova, S., Carretta, E., Randich, S., Tosi, M., 2008, A\&A,  480, 79
\bibitem[Bragaglia et al. (2012)]{bra12} Bragaglia, A., Gratton, R., Carretta, E., D'Orazi, V., Sneden, C., Lucatello, S.,2012, A\&A, 548, 122
\bibitem[Bressan et al. (2012)]{bre12} Bressan, A., Marigo, P., Girardi, L., Salasnich, B., Dal Cero, C., Rubele, S., Nanni, A., 2012, MNRAS, 427, 127
\bibitem[Carraro et al. (2002)]{car02} Carraro, G., Girardi, L., Marigo, P., 2002, MNRAS, 332, 705
\bibitem[Carraro et al. (2006)]{car06} Carraro, G., Subramaniam, A., Janes, K.A., 2006, MNRAS, 371, 1301
\bibitem[Carraro et al. (2008)]{car08} Carraro, G., V\'azquez, R.A., Moitinho, A., 2008, A\&A, 482, 777
\bibitem[Carraro and Seleznev (2011]{car11} Carraro, G., Seleznev, A.F., 2011, MNRAS, 412, 1361
\bibitem[Carraro (2013)]{car13} Carraro, G., 2013, proceedings of the 10th Pacific Rim Conference on Stellar Astrophysics,  ASP Conference Series, in press ({\tt 2013arXiv1308.5195C})
\bibitem[Carrera (2012)]{car12} Carrera, R., 2012, ApJ, 758, 110
\bibitem[Carretta e al. (2006)]{carr06} Carretta, E., Bragaglia, A., Gratton, R.G., Leone, F. Recio-Blanco, A., Lucatello, S.,2006, A\&A, 450, 523 
\bibitem[Carretta et al. (2014)]{car14} Carretta, E., Bragaglia, A., Gratton, R., D'Orazi, V.,  Lucatello, S., Sollima, A., 2014, A\&A,  561,87
\bibitem[Castelli et~al (1997)]{Castelli97}{Castelli} F., {Gratton} R.~G., {Kurucz} R.~L., 1997, \aap, 318, 841 
\bibitem[Coelho et al.(2005)]{coelho05} Coelho, P., Barbuy, B., Mel{\'e}ndez, J., Schiavon, R.~P., \& Castilho, B.~V.\ 2005, \aap, 443, 735 
\bibitem[Crawford (1958)]{cra58} Crawford, D.L., 1958, ApJ, 128, 185   
\bibitem[Dawson (1978)]{daw78} Dawson, D.W., 1978, AJ, 83, 1424
\bibitem[De Marchi et al. (2006)]{dem06} De Marchi, F., De Angeli, F., Piotto, G., Carraro, G., Davies, M.B., 2006, A\&A, 459, 489
\bibitem[Friel and Janes (1993)]{fri93} Friel, E.D., Janes, K.A., 1993, A\&A, 267, 75
\bibitem[Friel (1995)]{fri95} Friel, E.D., 1995, ARA\&A, 33, 381
\bibitem[Friel et al. (2002)]{fri02} Friel, E.D., Janes, K.A., Tavarez, M., Scott, J., Katsanis, R., Lotz, J., et al., 2002, AJ, 124, 2693
\bibitem[Geisler et al. (2012)]{gei12} Geisler, D., Villanova S., Carraro, G., Pilachowski, C., Cummings, J., Johnson, C.I., Bresolin, F., 2012, ApJ, 756, L40
\bibitem[Girardi et al. (2005)]{gir05} Girardi, L., Groenewegen, M.A.T., Hatziminaoglou,  E., da Costa, L., 2005, A\&A, 436, 895
\bibitem[Gratton \& Contarini (1994)]{gra94} Gratton, R., Contarini, G., 1994, A\&A, 283, 911
\bibitem[Grevesse and Sauval (1998)]{gs98} Grevesse, N., Sauval, A.~J., 1998, Space Science Reviews, 85, 161
\bibitem[Hamuy et al.(2006)]{ha06} Hamuy, M., Folatelli, G., Morrell, N.I., Phillips, M.M., et al., 2006, PASP, 118, 2
\bibitem[Hawarden et al.(1976)]{ha76} Hawarden, T.G., 1976, MNRAS, 174, 471
\bibitem[Hawarden et al.(1978)]{ha78} Hawarden, T.G., 1978, MNRAS, 182, 31 
\bibitem[Haywood et al. (2013)]{hay13} Haywood, M., Di Matteo, P., Lehnert, M.D>, Katz, D., G\'omez, A., 2013, A\&A, 560, 109
\bibitem[Hogg (1965)]{hog65} Hogg, A.R., 1965, Mem, Mt. Stromlo Obs., 17
\bibitem[Landolt (1992)]{la92} Landolt, A.U., 1992, AJ 104, 340
\bibitem[Kassis et al.(1997)]{ka97} Kassis, M., Janes, K.A., Friel, E.D., Phelps, R.A., 1997, AJ, 113, 1723
\bibitem[Keeping (1962)]{keeping62} Keeping, E., S., 1962. {\it Introduction to Statistical Inference} (Princeton: van Nostrand)
\bibitem[King (1964)]{ki64} King, I., 1964, R. Obs. Bull., No. 82
\bibitem[Kurucz(1993a)]{k93} Kurucz, R. L. 1993a, CD-ROM 13, 18~{\tt http://kurucz.harvard.edu}
\bibitem[Mashonkina et al.( 2007)]{ma07} Mashonkina, L. I., Vinogradova, A. B.,  Ptitsyn, D. A., Khokhlova, V. S.,
  Chernetsova, T. A., 2007, Astronomy Reports, 51, 903
\bibitem[Milone et al. (2009)]{mil09} Milone, A., Bedin, L.R., Piotto, G., Anderson, J., 2009, A\&A, 497, 755
\bibitem[Milone et al. (2012)]{mil12} Milone, A. P., Piotto, G., Bedin, L. R., et al. 2012, A\&A, 540, A16
\bibitem[Mitschang et al. (2014)]{mit14} Mitschang, A.W., De Silva, G., Zucker, D.B., Anguiano, B., Bensby, T., Feltzing, S., 2014, MNRAS, 438, 2753
\bibitem[Moitinho et al. (2006)]{moi06} Moitinho, A., V\'azquez, R.A., Carraro, G., Baume, G., Giorgi, E.E., Lyra, W., 2006, MNRAS, 368, L77
\bibitem[Patat and Carraro (2001)]{pa01} Patat, F., Carraro, G., 2001, MNRAS, 325, 1591
\bibitem[Schlegel et al. (1998)]{sc98} Schlegel, D.J., Finkbeiner, D.P., Davis, M., 1998, ApJ, 500, 525
\bibitem[Sestito et al. (2008)]{ses08} Sestito, P., Bragaglia, A., Randich, S., Pallavicini, R., Andrievsky, S.M., Korotin, S.A., 2008, A\&A, 488, 843
\bibitem[Sneden (1973)]{sneden73} Sneden, C. A., 1973, PhD thesis, THE UNIVERSITY OF TEXAS AT AUSTIN.
\bibitem[Sneden et al. (2003)]{sne03} Sneden, C.A., Preston, G.W., Cowan, J.J., 2003, ApJ, 592, 504
\bibitem[Sousa et~al. (2007)]{ares}Sousa S.~G., Santos N.~C., Israelian G., Mayor M., Monteiro M.~J.~P.~F.~G., 2007, A\&A, 469, 783
\bibitem[Stetson (1987)]{st87} Stetson, P.B., 1987, PASP 99, 191
\bibitem[Sloczewski et al. (2007)]{zl07} Zloczewski, K., Kaluzny, J., Krzemisnki, W., Olech, A., Thompson, I.B., 2007, MNRAS, 380, 1191
\bibitem[Villanova et al.(2013)]{vil13} Villanova, S., Geisler, D., Carraro, G., Moni Bidin, C., Munoz, C., 2013, ApJ, 778, 186
\end{thebibliography}
\end{document}